\documentclass[twocolumn,aps,pra,superscriptaddress,amsfonts,longbibliography]{revtex4-1}
\usepackage[colorlinks,linkcolor=blue,urlcolor=red,citecolor=red]{hyperref}
\usepackage{graphicx,amsmath,amsfonts,amssymb,capt-of}
\usepackage[tight]{subfigure}
\usepackage{siunitx}
\usepackage{bigints}
\usepackage{tabularx}
\usepackage{mathtools}
\usepackage{float}
\usepackage{multirow}
\usepackage{newfloat}
\usepackage{array}
\usepackage{tikz}
\usetikzlibrary{decorations.pathreplacing}
\usepackage{comment}
\usepackage{physics}
\usepackage[utf8]{inputenc}
\usepackage{amsmath}
\usepackage{geometry}
\usepackage{xcolor}
\usepackage{bbm}
\newcolumntype{P}[1]{>{\centering\arraybackslash}p{#1}}

\usepackage{enumitem}
\usepackage{chemfig}
\usepackage{circuitikz}
\usepackage{qcircuit}
\usepackage[normalem]{ulem}
\newcolumntype{x}[1]{>{\centering\let\newline\\\arraybackslash\hspace{0pt}}p{#1}}
\DeclareFloatingEnvironment[
    fileext=loa,
    listname=List of Algorithms,
    name=ALGORITHM,
    placement=tbhp,
]{algorithm}

\makeatletter
\newcommand{\vast}{\bBigg@{4}}
\newcommand{\Vast}{\bBigg@{5}}
\newcommand{\VAst}{\bBigg@{6}}
\makeatother

\usepackage{colortbl}
\begin{document}

\title{{Mapping  quantum chemical dynamics problems to spin-lattice simulators }}
\author{Debadrita Saha}
\author{Srinivasan S. Iyengar}
\email{Email: iyengar@indiana.edu}
\affiliation{Department of Chemistry, and the Indiana University Quantum Science and Engineering Center (IU-QSEC),
Indiana University, Bloomington, IN-47405}
\author{Philip Richerme}
\email{Email: richerme@indiana.edu }
\affiliation{Department of Physics and the Indiana University Quantum Science and Engineering Center (IU-QSEC),
Indiana University, Bloomington, IN-47405}
\author{Jeremy M. Smith}
\affiliation{Department of Chemistry,
Indiana University, 800 E. Kirkwood Ave, Bloomington, IN-47405}
\author{Amr Sabry}
\affiliation{Department of Computer Science, School of Informatics, Computing, and Engineering, and the Indiana University Quantum Science and Engineering Center (IU-QSEC),
Indiana University, Bloomington, IN-47405}
\date{\today}

\begin{abstract}
The accurate computational determination of  chemical, materials, biological, and atmospheric properties has critical impact on a wide range of health and environmental problems, but is deeply limited by the  computational scaling of quantum-mechanical methods.  The complexity of quantum-chemical studies arises from the steep algebraic scaling of electron correlation methods, and the exponential scaling in studying nuclear dynamics and molecular flexibility. To date, efforts to apply quantum hardware to such quantum chemistry problems have focused primarily on electron correlation. Here, we provide a framework which allows for the solution of quantum chemical nuclear dynamics by mapping these to quantum spin-lattice simulators. Using the example case of a short-strong hydrogen bonded system, we construct the Hamiltonian for the nuclear degrees of freedom on a single Born-Oppenheimer surface and show how it can be transformed to a generalized Ising model Hamiltonian. We then demonstrate a method to determine the local fields and spin-spin couplings needed to identically match the molecular and spin-lattice Hamiltonians.
We describe a protocol to determine the on-site and inter-site coupling parameters of this Ising Hamiltonian from the Born-Oppenheimer potential and nuclear kinetic energy operator. 
Our approach represents a paradigm shift in the methods used to study quantum nuclear dynamics, opening the possibility to solve both electronic structure and nuclear dynamics  problems using quantum computing systems.
\end{abstract}
\maketitle

\section{Introduction}
The quantum mechanical treatment of electrons and nuclei is critical for a wide range of problems that are of significance to biological, materials, and atmospheric studies. For example, hydrogen transfer processes  
are ubiquitous in reactions critical to human health, alternative energy sources, food security and environmental remediation \cite{Jeremy-23}. 
Yet, the detailed treatment of such problems is confounded by the presence of non-trivial quantum nuclear effects, such as hydrogen tunneling \cite{Klinman-Chemrevs-2006,SHS-PCET-SLO1,htrans,PCET-Nocera-ARPC-1998,PCET-Mayer-ARPC-1998}, coupled with electron correlation\cite{SHS-PCET}.
For the study of electron correlation in most molecular systems, 
several powerful approximations have been developed for classical computing platforms, 
and these are known to provide significant speedups compared to exponentially-scaling full-configuration interaction calculations. Indeed, chemical accuracy may be obtained for many systems using the well-known CCSD(T) method\cite{raghavachari1989fifth} that has an associated scaling cost of $\mathcal{O}(N^{6-7})$, where $N$ represents the number of electrons. 

More recently, algorithms to solve electron correlation problems in small molecular systems have been implemented on quantum hardware devices using trapped atomic ions, photons, nuclear spins, quantum dots, Rydberg atoms, and superconducting circuits\cite{o2016scalable,kandala2017hardware,xia2018quantum,gorman2018engineering,nam2019ground,Photosynthesis-Quantum-Sim,LHC-quantum-circuit,LHC-superconducting-circuits,peruzzo2014variational,grimsley2019adaptive,Google-12qubit-HF,Martinez-VQE-Solver-Excited-States}. 
The mapping of most electron correlation problems onto quantum hardware is facilitated by the Jordan-Wigner, parity, or Bravyi-Kitaev transformations \cite{Jordan-Wigner,Ortiz-JW,BRAVYI2002210}, where a product of Fermionic creation and annihilation operators are transformed to a chain of Pauli spin operators. 
In contrast, the intrinsic spin statistics of quantum nuclear dynamics problems do not play a role under conditions prevalent in biological, materials, and atmospheric systems, such as hydrogen transfer reactions under ambient conditions. As a result, most such quantum dynamics studies are currently constructed on classical computing platforms using basis sets and on grids.
Furthermore, many of these problems are known to display anomalous nuclear quantum effects\cite{Klinman-Chemrevs-2006,qwaimd-SLO-1,SLO-measurement} that are challenging to study 
due to the 
exponentially scaling computational cost of quantum dynamics with increasing degrees of freedom. 
Unlike several recent  attempts on the electron correlation problem\cite{o2016scalable,kandala2017hardware,xia2018quantum,gorman2018engineering,nam2019ground,Photosynthesis-Quantum-Sim,LHC-quantum-circuit,LHC-superconducting-circuits,peruzzo2014variational,grimsley2019adaptive,Google-12qubit-HF,Martinez-VQE-Solver-Excited-States,tkachenko2020correlation,cervera2020meta,huggins2021efficient,mcclean2020openfermion,motta2020quantum}, 
approximating quantum nuclear dynamics problems on quantum computing platforms has received relatively less attention\cite{kassal2008polynomial,macdonell2020analog,ollitrault2020hardware,sawaya2020resource,teplukhin2020solving,jahangiri2020quantum,wang2020efficient}. 
The primary goal of this paper is to develop a set of mapping protocols to allow the study of quantum nuclear dynamics problems on quantum hardware, without considering spin statistics. We provide and analyse 
an approximate algorithm 
to map exponentially-scaling quantum nuclear dynamics problems on a single Born-Oppenheimer surface, onto a general class of Ising-model Hamiltonians. Such Ising-type Hamiltonians may be implemented on a range of quantum computing platforms, such as ion-traps\cite{porras2004effective, richerme2014non}, super-conducting coils\cite{barends2014superconducting}, 
 Bosonic processors with photons\cite{Lanyon_Photonic-quantum-comp,aspuru2012photonic,knill2001scheme}, solid state devices and quantum dots inside cavities\cite{pellizzari1995decoherence,loss1998quantum,imamog1999quantum,PhysRevA.68.012310}, and Rydberg atoms\cite{saffman2010quantum,bernien2017probing}. Since quantum nuclear dynamics problems 
under ambient conditions do not need to be encoded using a set of Fermionic or Bosonic operators, we do not write the Ising model and molecular Hamiltonian in their respective second quantized forms. 
Instead, we first probe the structure of the Ising Hamiltonian matrix in its exponential scaling-space of spin basis vectors. This exponential space is admittedly intractable. 
Yet, our analysis of the Ising Hamiltonian matrix 
reveals an intrinsic structure where specific blocks appear within the Ising Hamiltonian matrix, and the corresponding matrix elements are only controlled by a subset of the externally controlled field parameters that dictate the dynamics of the model. To the best of our knowledge, such a structure has never been noted, or exploited, before in the literature. This structure allows us to characterize the general class of problems that may be ``computable'' using such hardware systems and in this paper we further inspect the extent to which quantum chemical dynamics studies may be conducted on such systems, when the statistics of particle permutation need not be included.

\begin{figure*}[tbp]
\includegraphics[width=\textwidth]{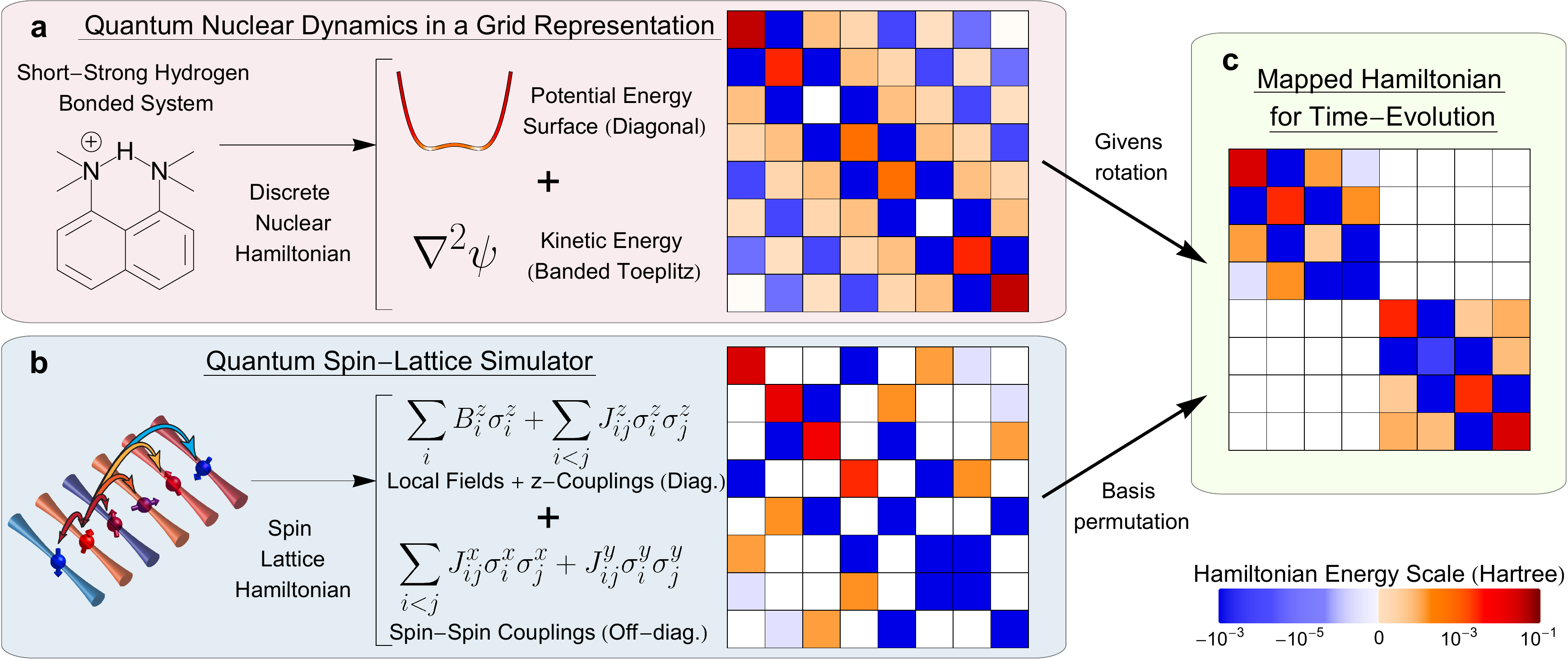}
\caption{\label{Map-outline}An outline of the mapping algorithm:
 The algorithm converts the Born-Oppenheimer potential surface and kinetic energy terms in a quantum-nuclear problem to a set of controllable parameters and facilitates the dynamical evolution of quantum states in an ion-trap. Box (a) shows the Born-Oppenheimer potential and kinetic energies for a short-strong hydrogen bonded system. This system Hamiltonian is mapped onto an ion trap quantum simulator shown in box (b).
Discrete representation of the nuclear Hamiltonian and appropriate rotations yield ion-trap parameters, $\left\{ \left\{ B_i^z \right\}; \left\{ J_{ij}^x, J_{ij}^y, J_{ij}^z \right\} \right\}$, 
 to determine the Ising model used to control the dynamics of lattice spin-states. }
\end{figure*}
The mapping algorithm 
is illustrated in Figure \ref{Map-outline}. 
An example of a quantum nuclear problem is shown in Figure \ref{Map-outline}a, where we depict a system containing a short-strong hydrogen bond with anharmonic vibrational behavior along the donor-acceptor axis. This problem is prototypical and is representative of a broad range of systems that occur during hydrogen transfer reactions\cite{Klinman-Chemrevs-2006} and in hydrogen-bonded systems that are known to have  significance in many critical processes\cite{Jeremy-23}.
We pre-compute the Born-Oppenheimer potential using electronic structure calculations and obtain a discrete version of the quantum nuclear Hamiltonian. 
To map this Hamiltonian onto a spin-lattice Ising-type model, the key insights in this paper are as follows: (i) A projected subspace of a specific unitary transformation of the diagonal elements of the quantum nuclear Hamiltonian (related to the Born-Oppenheimer potential) maps to and defines the local magnetic fields applied on each lattice site of an Ising model Hamiltonian. (ii) A similarly projected subspace of a related unitary transformation of the off-diagonal elements of the quantum nuclear Hamiltonian (related to the nuclear kinetic energy operator) defines and is mapped onto the inter-site coupling terms in the Ising model.  
Thus, we take a critical step towards solving quantum nuclear dynamics problems, and more generally problems that may not obey Fermi statistics, by mapping them to Ising-type Hamiltonians realizable on ion-trap quantum hardware. This is done without using a circuit model. The matrix elements, of the nuclear Hamiltonian that describe the molecular dynamics, inform the choice of local magnetic fields applied on each lattice site and laser pulse intensities that dictate the inter-site coupling, and govern the  dynamics of the ion-trap quantum computing platform. In this manner we provide a direct map of the two quantum systems. 

The paper is organized as follows: In Section \ref{Ising-Block}, we inspect the block structure of the Ising Hamiltonian which informs the general class of problems that may be computable on hardware architectures used to realize such Ising-type Hamiltonians. Following this, we then introduce the quantum nuclear Hamiltonian matrix on a single Born-Oppenheimer surface in Section \ref{Nuclear_Hamiltonian} and a class of Givens rotations\cite{golub2013matrix} based matrix transformations in Section \ref{Transformations_Hmol} to represent the quantum nuclear Hamiltonian matrix in a form that is commensurate with the transformed form of the Ising model Hamiltonian in Section \ref{Ising-Block}. This transformation leads to our approximate mapping protocol that is outlined in Section \ref{mapping}. Numerical results for the anharmonic molecular vibrations of the shared proton in a symmetric short-strong hydrogen bonded system are provided in Section \ref{results}. These include explicit numerical propagation of both the molecular dynamics problem as well as the spin lattice dynamics governed by Ising-type Hamiltonian where the Ising Hamiltonian parameters are chosen based on the mapping protocol in Section \ref{mapping}. The results match exactly for the case of three-qubits and error estimates beyond three-qubits are given in Section \ref{mapping}. Conclusions are given in Section \ref{concl}.

\section{Block structure of Ising-type Hamiltonian matrices obtained from appropriate classification of the computational basis}
\label{Ising-Block}
Ising-type Hamiltonians can be implemented on a range of available quantum computing platforms \cite{porras2004effective, richerme2014non,Lanyon_Photonic-quantum-comp,aspuru2012photonic,knill2001scheme,barends2014superconducting,pellizzari1995decoherence,loss1998quantum,imamog1999quantum,PhysRevA.68.012310,saffman2010quantum,bernien2017probing}, which makes these one of the most commonly-used quantum computing models today\cite{britton2012engineered,richerme2014non}. However, for specificity, we will illustrate our mapping protocols for ion-trap based quantum architectures, where ions form defect-free arrangements and can support quantum coherence times  longer than 10 minutes \cite{wang2017single}. Interactions between ions map to interactions between effective quantum spin states and quantum-harmonic-oscillator bath states -- each of which can be precisely controlled and programmed using laser light \cite{molmer1999multiparticle}. Site-resolved detection of each ion's spin state can be achieved with near-unit fidelity \cite{noek2013high}. These features have made trapped ions the leading platform for establishing atomic frequency standards \cite{ludlow2015optical} and one of the leading candidates for performing quantum simulations and quantum computations on such interacting spin systems. \cite{blatt2012quantum,islam2013emergence,richerme2013experimental,richerme2014non,senko2014coherent,smith2016many,zhang2017observation}

For ion-trap quantum hardware, the generalized Ising Hamiltonian is represented by a spin-lattice of qubits, where (a) the energy gap between the states at each qubit, $i$, and their relative orientations, are controlled by local effective magnetic fields, $\left\{ B_{i}^x, B_{i}^y, B_{i}^z \right\}$, and (b) the spin-spin coupling between different lattice sites, $i$ and $j$, is controlled using laser pulses, also spatially non-isotropic, and represented as $\left\{ J_{ij}^x, J_{ij}^y, J_{ij}^z \right\}$.
Thus, the most general Hamiltonian achievable within the ion trap quantum hardware at low temperatures is
\begin{eqnarray}
{\cal H}_{IT} &=& \sum_{\gamma}\sum_{\substack{i =1 \\ j>i}}  ^{N-1}J_{ij}^{\gamma}\sigma_{i}^{\gamma}\sigma_{j}^{\gamma} +\sum_{\gamma}\sum_{i=1}^{N}B_{i}^{\gamma} \sigma_{i}^{\gamma}
\label{HIT-gen}  
\end{eqnarray}
where $\gamma \in {(x,y,z)}$, and $N$ is the number of qubits (or ion-sites). 
The quantities $\left\{ \sigma_i^\gamma \right\}$ are the Pauli spin operators acting on the $i^{th}$ lattice site along the $\gamma$-direction of the Bloch sphere. 

In this paper, we map the Born-Oppenheimer nuclear Hamiltonian onto Eq. (\ref{HIT-gen}), thus allowing the two quantum systems to undergo analogous quantum dynamics. Towards this, the parameters 
$\left\{ B_{i}^\gamma; J_{ij}^\gamma \right\}$ 
are ``programmed'' as per the elements of the classically determined Born-Oppenheimer nuclear Hamiltonian matrix. To arrive at such a map, we first examine the intrinsic symmetries that are present within such  generalized Ising Hamiltonians.

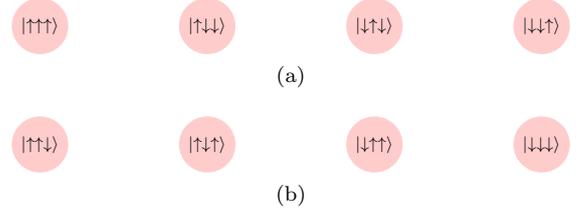
\begin{figure}[tbp]
\subfigure[]{
\begin{tikzpicture}
[scale=1.1,auto=center,every node/.style={circle,scale=0.7,fill=red!20}] 
     \node (a1) at (1,1) {$\ket{\uparrow \uparrow \uparrow}$};  
    \node (a2) at (3,1) {$\ket{\uparrow \downarrow \downarrow}$};    
    \node (a3) at (5,1) {$\ket{\downarrow \uparrow \downarrow}$};
    \node (a4) at (7,1) {$\ket{\downarrow \downarrow \uparrow}$};
\end{tikzpicture}
}
\hspace{2cm}
\subfigure[]{
\begin{tikzpicture}
[scale=1.1,auto=center,every node/.style={circle,scale=0.7,fill=red!20}] 
    \node (a7) at (9,1) {$\ket{\uparrow \uparrow \downarrow}$};  
    \node (a8) at (11,1) {$\ket{\uparrow \downarrow \uparrow}$};   
    \node (a9) at (13,1) {$\ket{\downarrow \uparrow \uparrow}$};
    \node (a10) at (15,1) {$\ket{\downarrow \downarrow \downarrow}$};
\end{tikzpicture}
}
\caption{\label{3qubit-blocks} The $2^N$ spin (computational) basis states are partitioned into spans of odd, $\left\{ {\bf S^+}^{2n-1} \right\}$ (Figure a) and even, $\left\{ {\bf S^+}^{2n} \right\}$ (Figure b) powers of the total spin raising operators. This is illustrated, here, for the case of three qubits. This leads to a block form of ${\cal H}_{IT}$ as illustrated in Figure \ref{3qubit-block-graph-main}.}
\end{figure}

\begin{figure*}[tbp]
\begin{tikzpicture}
[scale=1.1,auto=center,every node/.style={circle,scale=0.7,fill=red!20}]
     \node (a1) at (1,1) {$\ket{\uparrow \uparrow \uparrow}$};  
    \node (a2) at (3,1) {$\ket{\uparrow \downarrow \downarrow}$};    
    \node (a3) at (5,1) {$\ket{\downarrow \uparrow \downarrow}$};
    \node (a4) at (7,1) {$\ket{\downarrow \downarrow \uparrow}$};
    \node  (a5) at (2,2) {$J_{12}^{x}-J_{12}^{y}$};
    \node  (a6) at (6,2) {$J_{12}^{x}+J_{12}^{y}$};
    \node  (a13) at (4,2) {$J_{23}^{x}+J_{23}^{y}$};
    \node (a15) at (3,3) {$J_{13}^{x}-J_{13}^{y}$};
    \node (a16) at (5,3) {$J_{13}^{x}+J_{13}^{y}$};
    \node (a17) at (4,4) {$J_{23}^{x}-J_{23}^{y}$};
    \draw (a1) -- (a5);
    \draw (a5) -- (a15);
    \draw (a15) -- (a17);
    \draw (a2) -- (a5);
    \draw (a2) -- (a13);
    \draw (a3) -- (a13);
    \draw (a3) -- (a6);
    \draw (a6) -- (a4);
    \draw (a6) -- (a16);
    \draw (a13) -- (a15);
    \draw (a13) -- (a16);
    \draw (a16) -- (a17);
    \node (a7) at (9,1) {$\ket{\uparrow \uparrow \downarrow}$};  
    \node (a8) at (11,1) {$\ket{\uparrow \downarrow \uparrow}$};   
    \node (a9) at (13,1) {$\ket{\downarrow \uparrow \uparrow}$};
    \node (a10) at (15,1) {$\ket{\downarrow \downarrow \downarrow}$};
    \node (a11) at (10,2) {$J_{12}^{x}+J_{12}^{y}$};
    \node (a12) at (14,2) {$J_{12}^{x}-J_{12}^{y}$};
    \node (a14) at (12,2) {$J_{23}^{x}+J_{23}^{y}$};
    \node (a18) at (12,4) {$J_{23}^{x}-J_{23}^{y}$};
    \node (a19) at (11,3) {$J_{13}^{x}+J_{13}^{y}$};
    \node (a20) at (13,3) {$J_{13}^{x}-J_{13}^{y}$};
    \draw (a7) -- (a11);
    \draw (a8) -- (a11);
    \draw (a9) -- (a12);
    \draw (a10) -- (a12);
    \draw (a8) -- (a14);
    \draw (a9) -- (a14);
    \draw (a14) -- (a19);
    \draw (a14) -- (a20);
    \draw (a11) -- (a19);
    \draw (a12) -- (a20);
    \draw (a20) -- (a18);
    \draw (a19) -- (a18);
    \node (a21) at (8,2) {    };
    \node (a22) at (7,3) {$B_{3}^{x}+iB_{3}^{y}$};
    \node (a23) at (9,3) {$B_{3}^{x}+iB_{3}^{y}$};
    \node (a24) at (6,4) {$B_{2}^{x}+iB_{2}^{y}$};
    \node (a25) at (8,4) {   };
    \node (a26) at (10,4) {$B_{2}^{x}+iB_{2}^{y}$};
    \node (a27) at (5,5) {$B_{1}^{x}-iB_{1}^{y}$};
    \node (b1) [fill=gray!60] at (4,5) {  };
    \node (b2) [fill=gray!60] at (8,1) {  };
    \node (b3) [fill=gray!60] at (8,9) {  };
    \node (b4) [fill=gray!60] at (12,5) {  };
    \draw [gray!100] (b1) -- (b2);
    \draw [gray!100] (b2) -- (b4);
    \draw [gray!100] (b3) -- (b4);
    \draw [gray!100] (b3) -- (b1);
    \node (a28) at (7,5) {$B_{1}^{x}+iB_{1}^{y}$};
    \node (a29) at (9,5) {$B_{1}^{x}+iB_{1}^{y}$};
    \node (a30) at (11,5) {$B_{1}^{x}-iB_{1}^{y}$};
    \node (a31) at (6,6) {$B_{2}^{x}-iB_{2}^{y}$};
    \node (a32) at (8,6) {   };
    \node (a33) at (10,6) {$B_{2}^{x}-iB_{2}^{y}$};
    \node (a34) at (8,8) {    };
    \node (a35) at (7,7) {$B_{3}^{x}-iB_{3}^{y}$};
    \node (a36) at (9,7) {$B_{3}^{x}-iB_{3}^{y}$};
    \draw (a17) -- (a27);
    \draw (a16) -- (a24);
    \draw (a6) -- (a22);
    \draw (a4) -- (a21);
    \draw (a31) -- (a27);
    \draw (a28) -- (a24);
    \draw (a25) -- (a22);
    \draw (a23) -- (a21);
    \draw (a31) -- (a35);
    \draw (a28) -- (a32);
    \draw (a25) -- (a29);
    \draw (a23) -- (a26);
    \draw (a34) -- (a35);
    \draw (a36) -- (a32);
    \draw (a33) -- (a29);
    \draw (a30) -- (a26);
    \draw (a7) -- (a21);
    \draw (a11) -- (a23);
    \draw (a19) -- (a26);
    \draw (a18) -- (a30);
    \draw (a22) -- (a21);
    \draw (a25) -- (a23);
    \draw (a29) -- (a26);
    \draw (a33) -- (a30);
    \draw (a22) -- (a24);
    \draw (a25) -- (a28);
    \draw (a29) -- (a32);
    \draw (a33) -- (a36);
    \draw (a34) -- (a36);
    \draw (a27) -- (a24);
    \draw (a31) -- (a28);
    \draw (a35) -- (a32);
\end{tikzpicture}
\caption{\label{3qubit-block-graph-main} Block structure of ${\cal H}_{IT}$, Eq. (\ref{HIT-gen}), illustrated for a three qubit system: Computational (spin) basis state kets are presented at the base of the figure. These states are partitioned into odd, $\left\{ {\bf S^+}^{2n-1} \right\}$, and even, $\left\{ {\bf S^+}^{2n} \right\}$, spans of the total spin raising operators. The interaction between any two states,  $\ket{i}$ and $\ket{j}$ is the $ij^{th}$ matrix element of the ion trap Hamiltonian. For example, $\bra{\uparrow \downarrow \uparrow} {\cal H}_{IT} \ket{\downarrow \downarrow \downarrow} \equiv$ $\left[ J_{13}^{x} - J_{13}^{y} \right]$. The off-diagonal block that couples the vectors obtained from the odd, $\left\{ {\bf S^+}^{2n-1} \right\}$, and even, $\left\{ {\bf S^+}^{2n} \right\}$ spans of the total spin raising operators are marked using a gray square.}
\end{figure*}
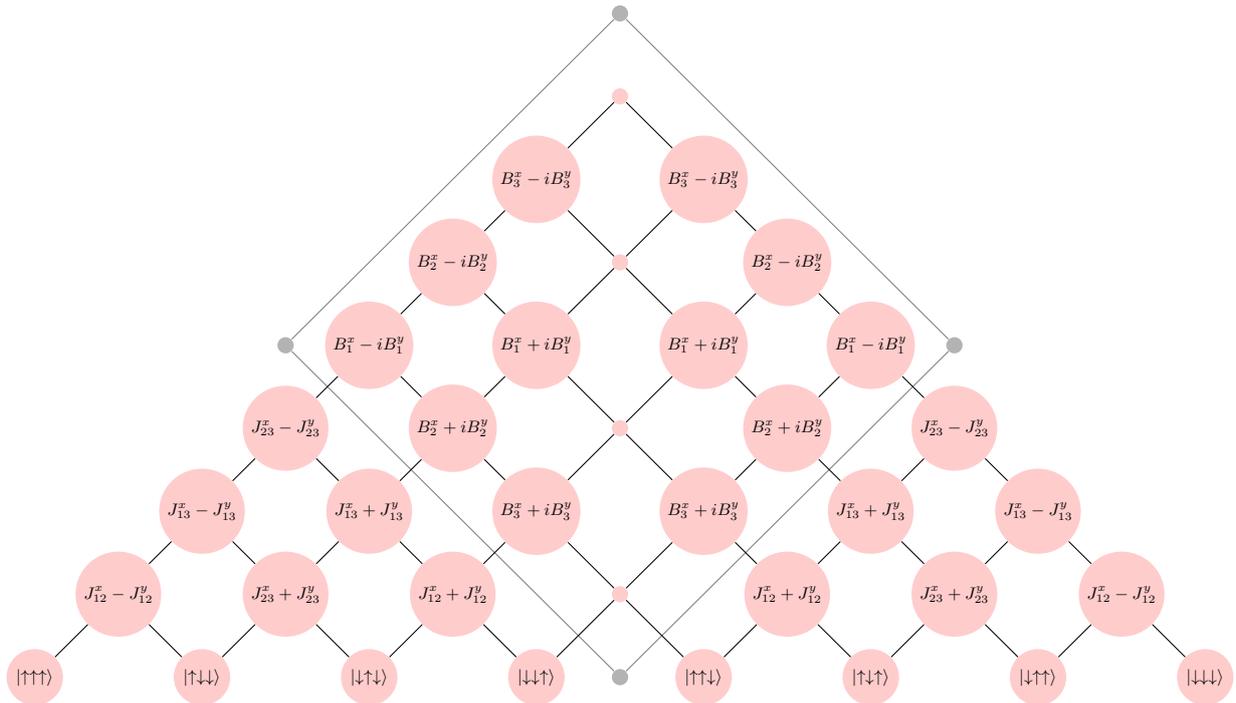

The ion-trap Hamiltonian, ${\cal H}_{IT}$, is naturally represented in a basis of $2^{N}$ spin states, where for example, $\left\{\ket{\uparrow \uparrow},  \ket{\uparrow \downarrow}, \ket{\downarrow \uparrow}, \ket{\downarrow \downarrow} \right\}$ form a basis for a 2-qubit system. These now provide us with a ``computational basis'' with programmable handles, $\left\{ B_{i}^\gamma; J_{ij}^\gamma \right\}$. To gauge the set of mappable problems, we introduce a general set of permutations on the 
computational basis vectors to reveal a novel block structure of the Ising Hamiltonian matrix. Specifically, the $2^{N}$ spin states are partitioned into two sets that are created by the span of even and odd total spin raising operators. Towards this, the basis vectors created from using an even number of lattice-site spin raising operators, $\left\{ S_i^+ \right\}$ acting on the full downspin state, $\ket{2^{N}-1} \equiv \ket{11\cdots} \equiv \ket{\downarrow \downarrow \cdots}$, yield the set, $\left\{ \ket{2^{N}-1};  S_i^+ S_j^+\ket{2^{N}-1} ; S_i^+ S_j^+ S_k^+ S_l^+\ket{2^{N}-1}; \cdots \right\}$, that are grouped as part of one block of the ion-trap Hamiltonian. See the set of vectors in 
Figure \ref{3qubit-blocks}a, and the bottom left row of Figure \ref{3qubit-block-graph-main}, where this idea is illustrated for a three-qubit system. For the notation in this paper, we have used the binary representation, $\ket{11\cdots}$ for spin state, $\ket{\downarrow \downarrow \cdots}$ and the corresponding integer representation for the state $\ket{2^{N}-1}$, obtained from the bit-sequence encoded in $\ket{11\cdots}$. 

Similarly, the states obtained using an odd number of raising operators:  $\left\{S_i^+\ket{2^{N}-1}; S_i^+ S_j^+ S_k^+\ket{2^{N}-1}; \cdots \right\}$ are grouped into a second block and are shown in Figure \ref{3qubit-blocks}b and on the bottom right row of Figure \ref{3qubit-block-graph-main}. Thus the two sets  independently span $\left\{ {\bf S^+}^{2n}  \ket{11\cdots} \right\}$ and $\left\{ {\bf S^+}^{2n-1}  \ket{11\cdots} \right\}$, where ${\bf S^+}$ is the total spin raising operator. When the spin basis vectors are partitioned in this fashion, the Ising Hamiltonian in Eq. (\ref{HIT-gen}) separates into the block structure that is illustrated in Figure \ref{3qubit-block-graph-main} for a three-qubit system. Specifically, the matrix that determines the time-evolution of the hardware system separates into two blocks that can only be coupled by turning on $\left\{ B_{i}^x; B_{i}^y \right\}$ and this is shown in Figure \ref{3qubit-block-graph-main} as part of the gray square. Thus eliminating these $\left\{ B_{i}^x; B_{i}^y \right\}$ fields would yield two separate blocks 
allowing the treatment of systems that may have a similar block structure. 
Similarly, the off-diagonal matrix elements within each block are determined by the laser field parameters, $\left\{ J_{ij}^x; J_{ij}^y \right\}$. While the structure derived here is completely general, it is illustrated in Figure \ref{3qubit-block-graph-main} for a 3-qubit system.  The diagonal elements of the matrix, not shown in Figure \ref{3qubit-block-graph-main} to maintain clarity, contain linear combinations of $\left\{ B_{i}^z;  J_{ij}^z \right\}$. 

For larger number of qubits, the block structure is recursive form and this aspect is further elaborated 
in Appendix \ref{Recursive_Structure_Ising}. This block-form of the Ising-type Hamiltonian and the associated structure in Figure \ref{3qubit-block-graph-main}, is a significant general result in this paper
, and as we find below, this analysis is critical towards mapping arbitrary problems. 

\section{The grid based quantum nuclear Hamiltonian computed on classical hardware}\label{Nuclear_Hamiltonian}
The quantum nuclear Hamiltonian for the molecular system, ${\cal H}^{Mol}$, is constructed on classical hardware, for the purpose of this paper. In the grid basis representation $\{ x\}$, the Hamiltonian matrix elements are given by,
\begin{equation}
    \bra{x} {{\cal H}}^{Mol} \ket{x^{\prime}} 
    = 
    K(x,x^{\prime}) + V(x)\delta_{x,x^{\prime}}
    \label{HDAF-molQD}
\end{equation}
For local potentials, the potential energy operator, $\hat{V}$, is diagonal in the
coordinate representation. The potential energy in the above equation is obtained from electronic structure calculations, that may also be performed on quantum hardware, independently, in future. 
The kinetic energy
may be approximated in a number of ways. One approach is to recognize that this operator is diagonal in the momentum representation and hence fast Fourier transforms are commonly employed 
\cite{feit1}. 
In this paper, we employ an analytic distributed approximating functional (DAF)
\cite{DAFprop-PRL,discreteDAF} representation for the coordinate
space version of 
the kinetic energy operator in Eq. (\ref{HDAF-molQD}): 
\begin{align}
K(x,x^{\prime}) =& K(\left\vert x-x^{\prime}\right\vert) =
\frac{-\hbar^2}{4m\sigma^3\sqrt{2\pi}}
\exp \left\{ -\frac{ {\left( x - x^\prime \right)}^2}
{2 {\sigma}^2} \right\} \nonumber \\ & \sum_{n=0}^{M_{DAF}/2} {\left( \frac{-1}{4} \right)}^n \frac{1}{n!} 
H_{2n+2} \left( \frac{ x - x^\prime }{ \sqrt{2} \sigma} \right).
\label{DAFfreeprop+derivative}
\end{align}
where,
$H_{2n+2}\left( \frac{ x - x^\prime }{ \sqrt{2} \sigma} \right)$ are the even order Hermite polynomials that only depend on the spread separating the grid basis vectors, $\ket{x}$ and $\ket{x^{\prime}}$, and $M_{DAF}$ and $\sigma$ are parameters that together determine the accuracy and efficiency of the resultant approximate kinetic energy operator. 
In this manner, the DAF presents a banded-Toeplitz representation for the kinetic energy operator, the structure of which, has a critical role in reducing the nuclear Hamiltonian to the form of ${\cal H}_{IT}$, depicted in Figure \ref{3qubit-block-graph-main}. This is further elaborated in the following section. 

\subsection{Unitary transformations that yield the Block structure of the nuclear Hamiltonian, for symmetric potentials, to make these commensurate with and mappable to the spin-lattice Hamiltonian, ${\cal H}_{IT}$}
\label{Transformations_Hmol}
The nuclear Hamiltonian, ${\cal H}^{Mol}$ from Eq. (\ref{HDAF-molQD}), has a banded Toeplitz structure due to the kinetic energy being expressed in terms of DAFs. In general, the Hamiltonian in Eq. (\ref{HDAF-molQD}) represents a multi-dimensional quantum dynamics problem, where the number of dimensions directly corresponds to the number of nuclear degrees of freedom. In this paper, we examine the map between the Hamiltonian in Eq. (\ref{HDAF-molQD}) for symmetric one-dimensional potentials and the Ising model Hamiltonian discussed in Section \ref{Ising-Block}. Routes from here to unsymmetric potentials and to problems in higher dimensions 
will be considered as part of future publications. In the proton transfer problem considered here, the one-dimensional potential energy surface along the hydrogen transfer axis, $V(x)$ in Eq. (\ref{HDAF-molQD}), is a symmetric double well owing to the isoenergetic donor and acceptor sites arising from the symmetry of the system (Figure \ref{Map-outline}(a)). 
We 
exploit the symmetric structure of the potential and the Toeplitz structure of the kinetic energy operator to construct a unitary transformation 
that block diagonalizes the nuclear Hamiltonian. 

\begin{figure*}[htb!]
    \centering
    \includegraphics[width=0.8\textwidth]{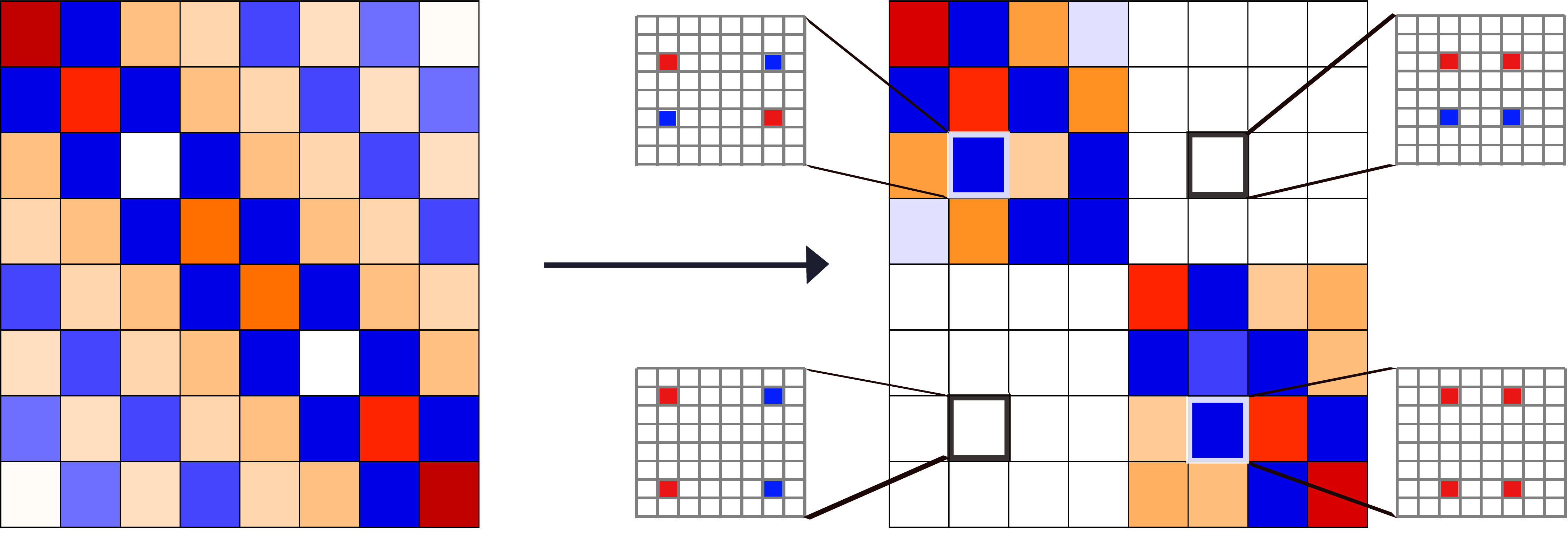}
    \caption{An illustration of the block-diagonalization  of the nuclear Hamiltonian, as captured by Eq. (\ref{Htilde-il}).  The original Hamiltonian, ${\cal H}^{Mol}$ is on the left, whereas the transformed $\Tilde{\cal{H}}^{Mol}$ is shown on the right. 
    On the right side, specific matrix elements from each block of  $\tilde{{\cal H}}_{il}^{Mol}$ are highlighted to illustrate Eqs. (\ref{Htilde-ii}) and (\ref{Htilde-il-od}). These highlighted elements of $\tilde{{\cal H}}_{il}^{Mol}$ are obtained by combining  elements of ${\cal H}^{Mol}$, as per Eq. (\ref{Htilde-il}), and these are marked using red and blue squares in zoomed in representations matrix elements in ${\cal H}^{Mol}$. The blue (negative) and red (positive) indicate the phase of the corresponding elements of ${\cal H}^{Mol}$, as obtained from $\alpha_{i}$ in Eqs. (\ref{Htilde-il}), (\ref{Htilde-ii}) and (\ref{Htilde-il-od}). 
    }
    \label{Unitary_transformH}
\end{figure*}
The unitary transform that leads to the block diagonalization of the nuclear Hamiltonian, similar to the structure of the Ising Hamiltonian, can be expressed as a product of Givens rotations. The effect of the Givens rotations on the grid basis states is to create superposition states of the symmetric grid basis states. The action of the product of Givens rotations is, therefore, to yield a rotation in the $2^{N}$ dimensional basis state space, by a sequence of 2$\times$2 (or one-qubit) rotations. 
This now divides the basis state space, in which the Hamiltonian is represented, into two sets of
rotated states, 
created by the symmetric and anti-symmetric combinations respectively. To explain this, we introduce a uniform one-dimensional set of $2^{N}$ grid points, $\left\{ x_i \right\}$, such that the Givens transformed grid basis, $\left\{ \tilde{x}_i \right\}$, may be represented as 
 \begin{align}
  \ket{\tilde{x}_{i}} \equiv \frac{1}{\sqrt{2}}\left[{\ket{x_{i}} \pm \ket{x_{2^{N}+1-i}}}\right], \quad 1 \leq i \leq 2^{N}
  \label{Givens-transformed-basis}
 \end{align}
These now form two mutually orthogonal subspaces that block diagonalize the nuclear Hamiltonian for symmetric potentials. This process is illustrated for a three-qubit system ($2^3$-grid points) in Figure \ref{Unitary_transformH}.
The $il^{th}$ matrix element of the resultant molecular Hamiltonian in the Givens transformed grid basis 
is explicitly written as
  \begin{align}
      \tilde{\mathcal{H}}^{Mol}_{il} = \frac{1}{2}&\left(\mathcal{H}^{Mol}_{i,l} + \alpha_{l}\mathcal{H}^{Mol}_{i,n+1-l} + \alpha_{i}\mathcal{H}^{Mol}_{n+1-i,l} + \right. \nonumber \\ & \left. \hphantom{(}\alpha_{i}\alpha_{l}\mathcal{H}^{Mol}_{n+1-i,n+1-l}\right),
      \label{Htilde-il}
  \end{align}
where $n= 2^N$ and $\alpha_{i}=\text{sgn}\left[i-(n+1)/2\right]$. 
The elements of the diagonal blocks of $\tilde{\mathcal{H}}^{Mol}$ (matrix on the right in Figure (\ref{Unitary_transformH})) are obtained from Eq. (\ref{Htilde-il}) as 
   \begin{align}
      \tilde{\mathcal{H}}^{Mol}_{il} &= \frac{1}{2}\left(\mathcal{H}^{Mol}_{i,l} + \alpha_{i}\mathcal{H}^{Mol}_{i,n+1-l} + \alpha_{i}\mathcal{H}^{Mol}_{n+1-i,l} + \right. \nonumber \\ & \left. \hphantom{= \frac{1}{2}(} \mathcal{H}^{Mol}_{n+1-i,n+1-l}\right) \nonumber \\ &= \left[ {K}(x_{i},x_{l}) + \alpha_{i} {K}(x_{i},x_{n+1-l}) \right] + \nonumber \\ &  \hphantom{= \frac{1}{2}(}\frac{1}{2} \left[ V(x_{i}) + V(x_{n+1-l}) \right] \delta_{i,l}
      \label{Htilde-ii}
  \end{align}
The elements of the unitary transform, $\alpha_{i}$ are, in fact, the characters of the $C_s$ point group. The right hand side of the above equation, therefore, represents a symmetry adapted transformation of the nuclear Hamiltonian, and the term $\frac{1}{2} \left[ V(x_{i}) + V(x_{n+1-i}) \right]$, symmeterizes the potential energy surface in one-dimension. 
By extension, for the off-diagonal blocks of $\tilde{\mathcal{H}}^{Mol}$ in Figure \ref{Unitary_transformH}, $\alpha_{l} = -\alpha_{i}$ and   \begin{align}
      \tilde{\mathcal{H}}^{Mol}_{il} =& \frac{1}{2}\left(\mathcal{H}^{Mol}_{i,l} - \alpha_{i}\mathcal{H}^{Mol}_{i,n+1-l} + \alpha_{i}\mathcal{H}^{Mol}_{n+1-i,l} - \right. \nonumber \\ & \left. \hphantom{\frac{1}{2}(}\mathcal{H}^{Mol}_{n+1-i,n+1-l}\right) \nonumber \\ =& 
      \frac{1}{2} \left[ V(x_{i}) - V(x_{n+1-l}) \right]  
      \delta_{i,n+1-l}
      \label{Htilde-il-od}
  \end{align}
where the kinetic energy contribution is  identically zero purely due to the Toeplitz nature of Eq. (\ref{DAFfreeprop+derivative}), and only the anti-symmetric portion of the potential, $\frac{1}{2} \left[ V(x_{i}) - V(x_{n+1-l}) \right]$, contributes to the anti-diagonal part of $\tilde{\mathcal{H}}^{Mol}$. Thus for symmetric potentials such as those considered here, Eq. (\ref{Htilde-il-od}) is identically zero. This observation will become useful when we generalize the approach presented here, first to general potentials and then, to problems of higher dimensionality in future publications.

\section{Mapping protocol for quantum chemical dynamics} 
\label{mapping}
The structure of the ion-trap Hamiltonian constrains the class of mappable problems. 
These constraints dictate the accuracy with which quantum chemical dynamics simulations can be performed on an ion-trap system given by Eq. (\ref{HIT-gen}). 
To summarize our discussion thus far, we began with a computational basis $\ket{\lambda}$, used to describe the Ising Hamiltonian, ${\cal H}_{IT}$ in Eq. (\ref{HIT-gen}), and the molecular basis $\ket{x}$, used to represent the quantum nuclear Hamiltonian, ${\cal H}^{Mol}$ in Eq. (\ref{HDAF-molQD}). In the interest of matching the structures of the two Hamiltonians, we first obtained a permuted computational basis: $\ket{\lambda} \rightarrow \ket{\tilde{\lambda}}$ (Section \ref{Ising-Block}) and a unitary (Givens) transformed quantum nuclear basis: $\ket{x} \rightarrow \ket{\tilde{x}}$ (Section \ref{Transformations_Hmol}). In doing so our goal becomes: 
\begin{align}
       {\bra{\tilde{x}}} \mathcal{H}^{Mol} {\ket{\tilde{x^\prime}}} \leftrightarrow {\bra{\tilde{\lambda}}} {\cal H}_{\text {IT}} {\ket{\tilde{\lambda^\prime}}}
       \label{HMol-HIT-correspondence}
\end{align} 
where we have also tersely introduced a map between the transformed quantum nuclear wavefunction bases and the permuted computational bases that represent the Ising spin lattice system as:
\begin{align}
    \ket{\tilde{x}} \Leftrightarrow{}
    \ket{\tilde{\lambda}}.
    \label{lambda-x-correspondence}
\end{align}
The effectiveness of the maps in Eqs. (\ref{HMol-HIT-correspondence}) and (\ref{lambda-x-correspondence}) will essentially dictate the accuracy to which the dynamics captured within the ion-trap quantum simulator controlled by an Ising Hamiltonian accurately predicts the quantum nuclear dynamics. 
In this section, we will show that, due to the structure of the Hamiltonians discussed in the previous sections, the diagonal and off-diagonal elements of each individual diagonal block of mappable Hamiltonians, such as Eq. (\ref{HDAF-molQD}), are Hadamard transformed to provide $\left\{ B_i^z;  J_{ij}^z \right\}$ and $\left\{ J_{ij}^x; J_{ij}^y \right\}$, respectively. 
Similarly the coupling between the diagonal blocks 
in the Ising Hamiltonian, are controlled through $\left\{ B_{i}^x; B_{i}^y \right\}$ and provide handles for the appropriate blocks of mappable Hamiltonians. As a consequence of the discussion in Section \ref{Transformations_Hmol}, both Hamiltonians, Eqs. (\ref{HIT-gen}) and (\ref{HDAF-molQD}), by construction, take the form depicted in Figure \ref{Map-outline}c and the right side of Figure \ref{Unitary_transformH}, respectively. However, we will also see that these maps are not exact for arbitrary Hamiltonians, beyond three-qubits, and 
towards the end of this section we provide error-bounds to determine the extent to which the Ising Hamiltonians deviate from the molecular Hamiltonian (or any other general matrix). 
 Our quantum nuclear dynamics test case that will be mapped to the aforementioned Ising Hamiltonian (Section \ref{results}), 
exploits the block structure discussed above and we illustrate the map by studying a symmetric hydrogen bonded system displayed in Figure \ref{Map-outline}a, where a symmetric double-well potential is also shown. For these cases, as seen from Eqs. (\ref{Htilde-il-od}), (\ref{Htilde-ii}) and the discussion in Section \ref{Ising-Block} and Eq. (\ref{HIT-nqubit-Block}) in Appendix \ref{Recursive_Structure_Ising}, the block structure of both the Hamiltonian matrices allows the two blocks of each Hamiltonian matrix (Ising and molecular) to be propagated independently, and, potentially on different quantum simulators, for the Ising Hamiltonian. We exploit this feature to evaluate a separate set of $\left\{ B_i^z;  J_{ij}^{\gamma} \right\}$ values, below, for each of the two diagonal blocks of the molecular Hamiltonian, while maintaining $\left\{ B_{i}^x; B_{i}^y \right\}$ to be identically zero.  

\subsection {Obtaining ion-trap parameters $\left\{ B_{i}^z;  J_{ij}^z \right\}$ from 
the  diagonal elements of the molecular Hamiltonian }{\label{map-diag}}
The diagonal elements of the molecular Hamiltonian are directly mapped to those of the spin lattice Hamiltonian after invoking the map of the unitary transformed grid basis ($\ket{\tilde{x}}$) to the permuted computational basis ($\ket{\tilde{\lambda}}$). Each diagonal element of the molecular Hamiltonian in the transformed grid representation, ${\bra{\tilde{x}}}\mathcal{H}^{Mol} {\ket{\tilde{x}}}$, is equivalent to the corresponding element of the ion-trap Hamiltonian, ${\bra{\tilde{\lambda}}} {\cal H}_{\text {IT}} {\ket{\tilde{\lambda}}}$ in the permuted computational basis representation. See Eqs. (6) and (7). In doing so, the set of on-site and inter-site coupling parameters, $\left\{ B_i^z;  J_{ij}^z \right\}$, of the ion-trap that occur along the diagonal of $\tilde{{\cal H}}_{\text {IT}}$ can be evaluated.
The mapping expression between the diagonal elements of the molecular Hamiltonian and the ion-trap Hamiltonian may be written as
    \begin{align}
       {\bra{\tilde{x}}} \mathcal{H}^{Mol} {\ket{\tilde{x}}} \equiv& {\bra{\tilde{\lambda}}} {\cal H}_{\text {IT}} {\ket{\tilde{\lambda}}} = \nonumber \\ & \sum_{j=1}^{N} (-1)^{\tilde{\lambda}_{j}} B_{j}^{z} + \sum_{j=1}^{N-1} \sum_{k>j}^{N} (-1)^{\tilde{\lambda}_{j}\oplus\tilde{\lambda}_{k}} J_{jk}^{z}
        \label{Diag-B}
    \end{align} 
    where $\oplus$ denotes the addition modulo 2, 
    $\tilde{\lambda}_{j}$ is the $j^{th}$ bit of the bit representation of $\ket{\tilde{\lambda}}$ with values $0$ or $1$ for up- or down-spin, respectively, as shown in Figures \ref{3qubit-blocks} and \ref{3qubit-block-graph-main}. 
    Our goal here now is to be able to simulate the diagonal elements of $\tilde{\mathcal{H}}^{Mol}$ that are influenced by the Born-Oppenheimer potential, by tuning the set of ion-trap parameters $\{B^{z}_{i},J^{z}_{ij}\}$ that determine the diagonal elements of
    $\tilde{{\cal H}}_{\text {IT}}$. 
    
To this effect, we treat all parameters in $\left\{ B_i^z;  J_{ij}^z \right\}$ on an equal footing and write the mapping expression for the diagonal elements in each block as,  
    \begin{align}
        {\bra{\tilde{x}}} \mathcal{H}^{Mol} {\ket{\tilde{x}}} \equiv {\bra{\tilde{\lambda}}} {\cal H}_{\text {IT}} {\ket{\tilde{\lambda}}} = \sum_{i=1}^{\frac{N(N+1)}{2}} {\bf T}_{\tilde{\lambda},i} D_{i}^{z}
        \label{Diag-map}
    \end{align}
    where $\{\tilde{\lambda}\}$ correspond to either of the two sets of permuted computational basis states (see Figure \ref{3qubit-blocks}) that independently span $\left\{ {\bf S^+}^{2n}  \ket{11\cdots} \right\}$ or $\left\{ {\bf S^+}^{2n-1}  \ket{11\cdots} \right\}$, $D^{z}$ is a vector that represents in a consolidated fashion, the individual $\left\{ B_{j}^{z};  J_{jk}^{z}\right\}$ parameters 
    and the ${\bf T}_{\tilde{\lambda},i}$ represents a coefficient matrix for the phase preceding the corresponding $D_{i}^{z}$s as seen from Eq. (\ref{Diag-B}). The upperlimit for the index $i$ in the above expression thus denotes the maximum number of independent parameters in $\left\{ B_{j}^{z};  J_{jk}^{z}\right\}$ that will be used to encode the diagonal part of $\tilde{\mathcal{H}}^{Mol}$ for a given number of qubits. This aspect of matching the number of parameters is further discussed in Appendix \ref{DoF_Ising}.  
    Thus, the diagonal elements of the 
    transformed 
    Ising Hamiltonian, ${\bra{\tilde{\lambda}}} {\cal H}_{\text {IT}} {\ket{\tilde{\lambda}}}$ encode 
    both the Born-Oppenheimer potential energy surface, $V(x)$, 
    and the 
    Givens transformed elements of the 
    nuclear kinetic energy 
    that appear due to block diagonalization process, needed to make the two Hamiltonians have the same structure (See Eq.(\ref{Htilde-ii})). 
    The columns of the transformation matrix on the right side of Eq. (\ref{Diag-map}) with elements, 
   ${\bf T}_{\tilde{\lambda},i} = \pm 1$, resemble a subset of columns that span a $N-1$-dimensional Hadamard matrix, which is a $(N-1)^{th}$ order tensor product ($H^{\otimes (N-1)}$) 
   of the standard 2$\times$2 (or 1-qubit) Hadamard transform $H$. 
   Thus, the columns of ${\bf T}$ in Eq. (\ref{Diag-map}) are orthogonal and the 
   $\left\{ B_{j}^{z};  J_{jk}^{z}\right\}$ values may be computed as: 
    \begin{equation}
        D^{z}_{j} = \frac{1}{2^{N-1}}\sum_{\{\tilde{\lambda}\}} \mathbf{T}_{j,\tilde{\lambda}}{\bra{\tilde{\lambda}}} {\cal H}_{\text {IT}} {\ket{\tilde{\lambda}}}. 
        \label{Diag-B-map1}
    \end{equation}
\begin{figure*}
        \begin{minipage}{0.4\textwidth}
        \subfigure[3-qubits: $ {\bra{\tilde{x}}} \mathcal{H}^{Mol} {\ket{\tilde{x}}} \rightarrow B_{j}^{z}$]{\includegraphics[width=\textwidth]{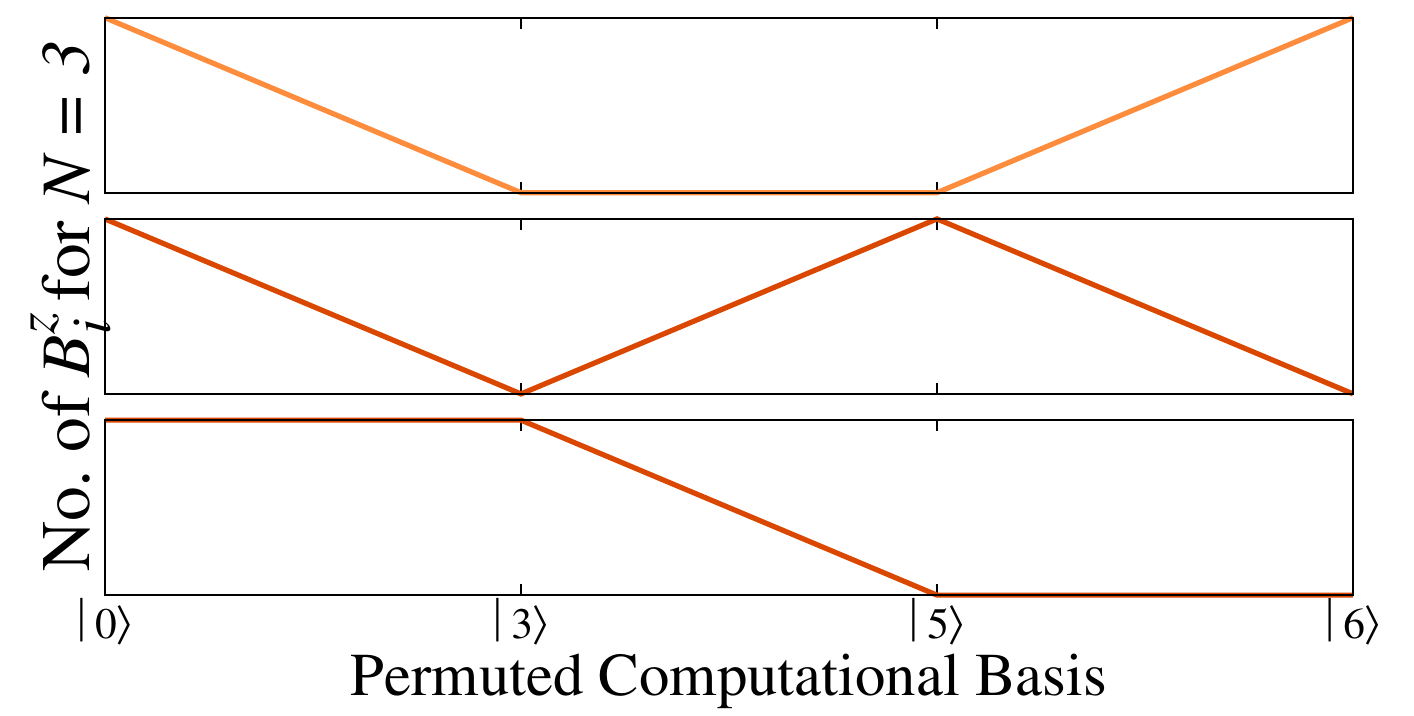}}
        \subfigure[5-qubits: $ {\bra{\tilde{x}}} \mathcal{H}^{Mol} {\ket{\tilde{x}}} \rightarrow B_{j}^{z}$]{\includegraphics[width=\textwidth]{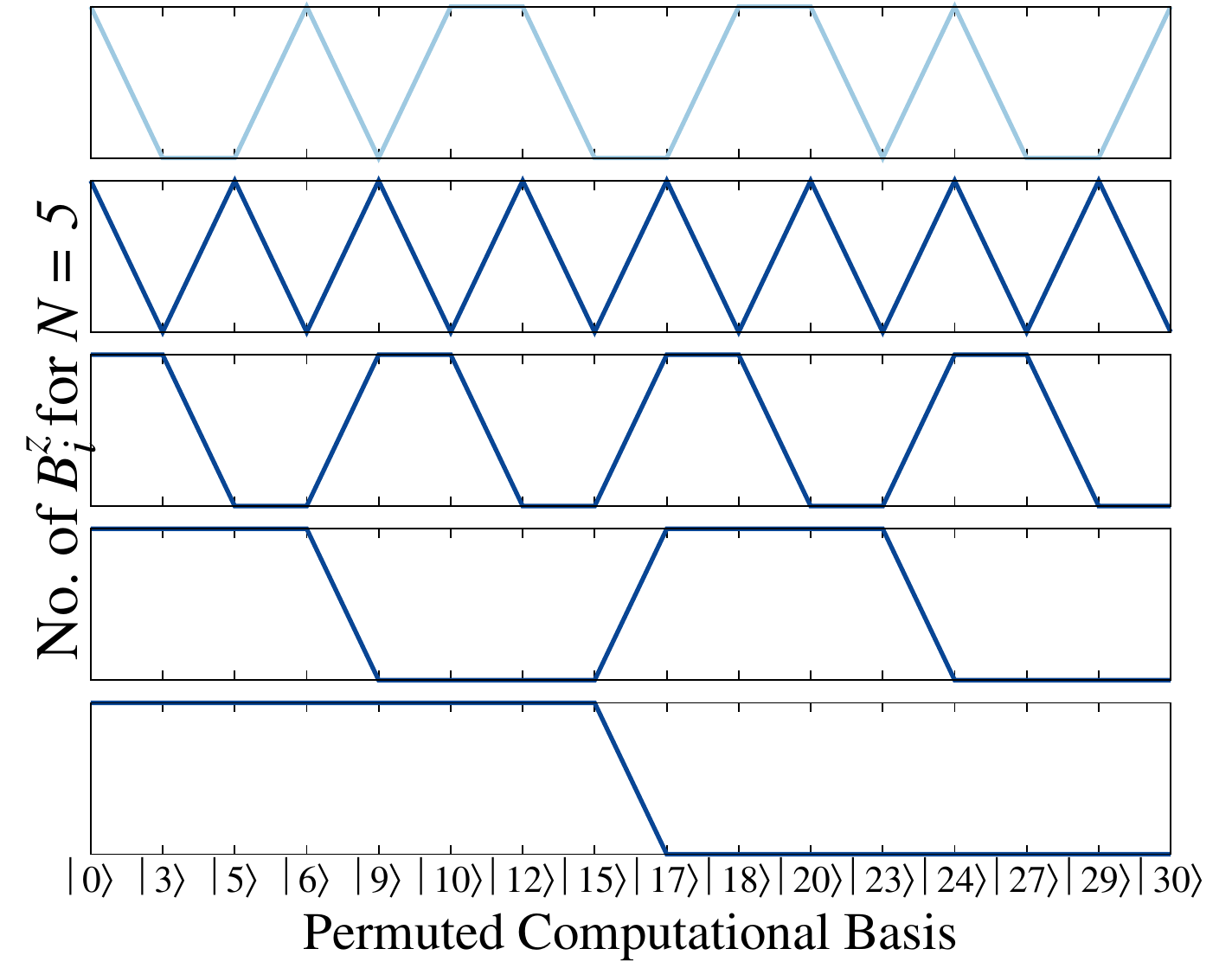}}
        \end{minipage}
        \begin{minipage}{0.4\textwidth}
        \subfigure[5-qubits: $ {\bra{\tilde{x}}} \mathcal{H}^{Mol} {\ket{\tilde{x}}} \rightarrow J_{jk}^{z}$]{\includegraphics[width=\textwidth]{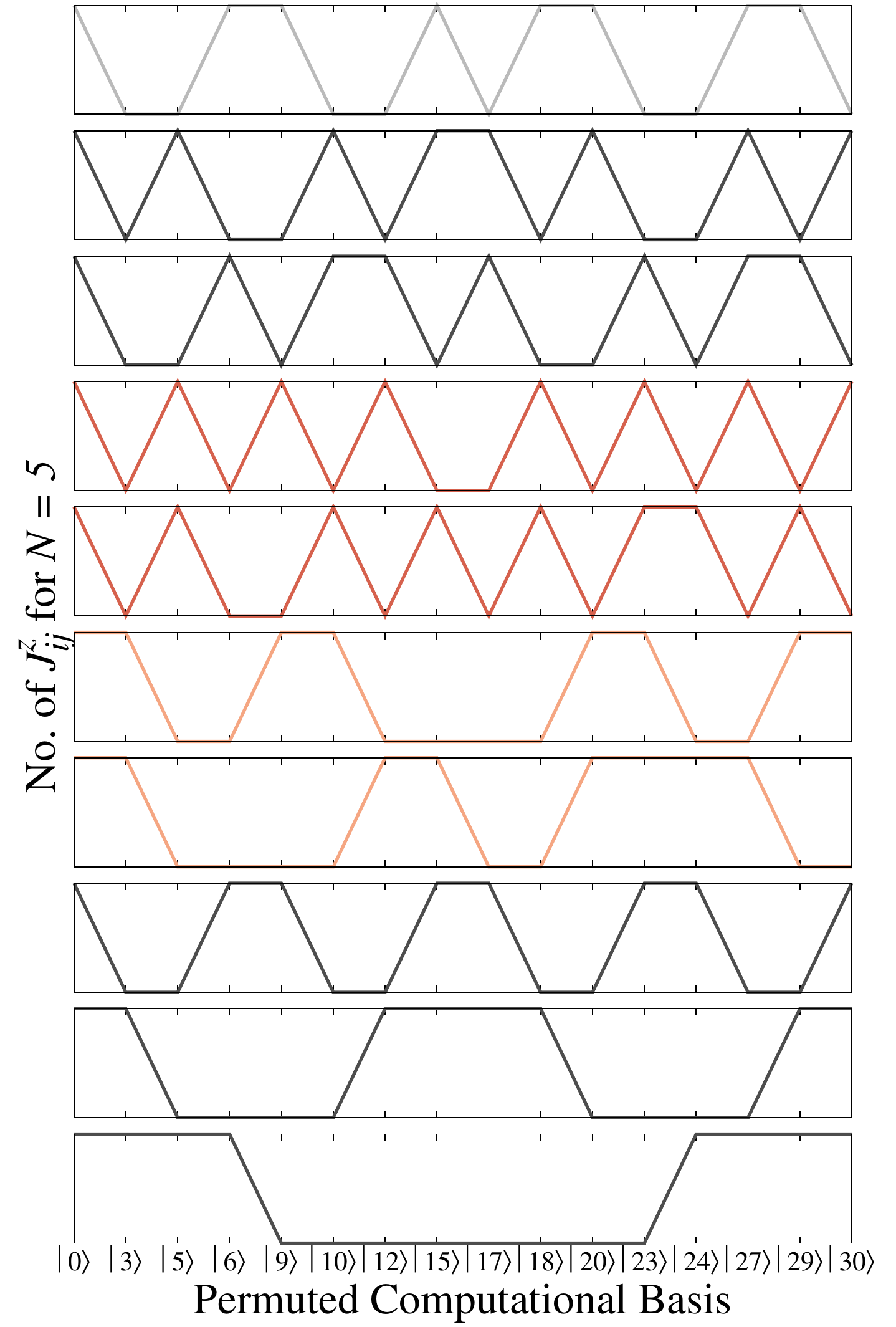}}
        \end{minipage}
       \caption{\label{fig:T-VtoBz} Figure illustrates the  transformation from $ {\bra{\tilde{x}}} \mathcal{H}^{Mol} {\ket{\tilde{x}}}$ to the $B^{z}$-values for 3-qubits (a) and $B^{z}$ (b) and $J^{z}$-values (c) for 5-qubits. 
       This inverse matrix picks out the appropriate columns from the $2^{N-1}$-dimensional Hadamard transformation matrix. 
       See Eqs. (\ref{Diag-B}) and (\ref{Diag-B-map2}). 
       The $B^{z}$ transformations (Figures a and b) arise from two basic vectors, (top panel and bottom panel). The vector in bottom panel is scaled in frequency to create the other vectors. This aspect is show by using two different color shades. Similarly,  $J^{z}$ transformations (Figure c) arise from five basic transformations arranged with different color shades in panels 1-3, 4-5, 6-7, 8-9 and 10.
       }
\end{figure*}
    Owing to the equivalence of ${\bra{\tilde{x}}} \mathcal{H}^{Mol} {\ket{\tilde{x}}}$ and ${\bra{\tilde{\lambda}}} {\cal H}_{\text {IT}} {\ket{\tilde{\lambda}}}$, as seen in Eq. (\ref{HMol-HIT-correspondence}), we use the precomputed diagonal elements of the unitary transformed molecular Hamiltonian in Eq.(\ref{Diag-B-map1}) to obtain the on-site parameters of the ion-trap. As per Eq. (\ref{Diag-B}), ${\bra{\tilde{\lambda}}} {\cal H}_{\text {IT}} {\ket{\tilde{\lambda}}}$ may be replaced by ${\bra{\tilde{x}}} \mathcal{H}^{Mol} {\ket{\tilde{x}}}$ in Eq. (\ref{Diag-B-map1}) leading to, 
    \begin{equation}
        D^{z}_{j} = \frac{1}{2^{N-1}} \sum_{\{\tilde{x}\}\Leftrightarrow{}\{\tilde{\lambda}\}} \mathbf{T}_{j,\tilde{\lambda}}  {\bra{\tilde{x}}} \mathcal{H}^{Mol} {\ket{\tilde{x}}} 
        \label{Diag-B-map2}
    \end{equation}

\noindent where we tersely assume the summation over $\{\tilde{\lambda}\}$ to also correspond to the summation over $\{\tilde{x}\}$ 
as allowed by the correspondence in Eq. (\ref{lambda-x-correspondence}). 
The transformation matrix in Eq. (\ref{Diag-B-map2}), that is $\mathbf{T}^{T}$, is illustrated in Figure \ref{fig:T-VtoBz} (b and c) for the $2^{4}$-dimensional sub-blocks of a 5-qubit Ising and for the $2^{2}$-dimensional sub-blocks of a 3-qubit Ising Hamiltonian in Figure \ref{fig:T-VtoBz} (a). 
The dimension, $2^{N-1}\times\frac{N(N+1)}{2}$ of the $\mathbf{T}$ matrix is apparent from this figure. The latter dimension of the $\mathbf{T}$ matrix that depends on the number of independent $D^{z}_{j}$ values is at most $\frac{N(N+1)}{2}$, and is found to be 3 for the 3-qubit system and 15 for the 5-qubit system. More general expressions comparing the number of Ising Hamiltonian control parameters to the number of independent matrix elements in the molecular Hamiltonian, are given in Appendix \ref{DoF_Ising}. 
While the figure is only presented for 3-qubit and 5-qubit systems, the transformation is completely general.
    
At this stage, as noted in 
Appendix \ref{DoF_Ising}, Eqs. (\ref{IT-handles1}) and (\ref{IT-handles2}), and by comparison with the upper limit of the summation in Eq. (\ref{Diag-map}), where clearly $\frac{N(N+1)}{2} < 2^{N-1}$ for $N>4$, it is clear that the number of $\left\{ B_i^z;  J_{ij}^z \right\}$ ion-trap handles are fewer than the number of diagonal elements in the molecular Hamiltonian for large number of qubits. However, it may be possible to pre-rotate the molecular Hamiltonian basis (similar to the Givens rotation based unitary transform in Section (\ref{Transformations_Hmol})) so as to compress the amount of information along the diagonal and these aspects will be considered in a future publication.

\subsubsection{Error bounds on mapping ${\bra{\tilde{x}}} \mathcal{H}^{Mol} {\ket{\tilde{x}}} \leftrightarrow {\bra{\tilde{\lambda}}} {\cal H}_{\text {IT}} {\ket{\tilde{\lambda}}}$ for larger number of qubits}\label{Error_bounds_section} 
Arising from the above discussion, the error, $\mathbf{\epsilon}$, associated with such a partial Hadamard transform of the diagonal elements of the molecular Hamiltonian, can be obtained from the orthogonal complement of the transformation matrix in the corresponding Hadamard matrix. This can be expressed in a closed form as,
    \begin{equation}
        \mathbf{\epsilon} = \frac{1}{2^{N-1}}\sqrt{{([\tilde{\mathcal{H}}^{Mol}_{Diag}]^{T}\mathbf{P^{T_{\perp}}} [\tilde{\mathcal{H}}^{Mol}_{Diag}]}}
        \label{error_Bz}
    \end{equation}
    where $\tilde{\mathcal{H}}^{Mol}_{Diag}$ contains the diagonal elements of $\tilde{\mathcal{H}}^{Mol}$, that is ${\bra{\tilde{x}}} \mathcal{H}^{Mol} {\ket{\tilde{x}}}$ in the equations above and $\mathbf{P^{T_{\perp}}}$ is a projector on to the orthogonal complement of transformation matrix ${\bf T}$ as obtained from the Hadamard matrix.
    \begin{align}
        \mathbf{P^{T_{\perp}}} &\equiv 2^{N-1}H^{\otimes (N-1)}{H^{\otimes (N-1)}}^T - {\bf T}{\bf T}^T \nonumber \\ &= 2^{N-1}\mathbf{I_{2^{(N-1)}}} - {\bf T}{\bf T}^T
    \end{align}
 Thus, in cases where the diagonal part of the molecular Hamiltonian are exactly captured within the subspace represented by Eq. (\ref{Diag-B-map2}), may be exactly modeled using the ion-trap simulator/computer. In all these cases the orthogonal complement in Eq. (\ref{error_Bz}) is identically zero. An illustration of the transformation matrices that are used to compute $\left\{ B_i^z;  J_{ij}^z \right\}$ and the errors corresponding to the simulation of the diagonal elements, for the particular case of three qubits, is provided in Appendix \ref{Diagonal-transform-3qubit}.

\begin{figure}[bp]
\subfigure[]{\includegraphics[height=0.4\columnwidth]{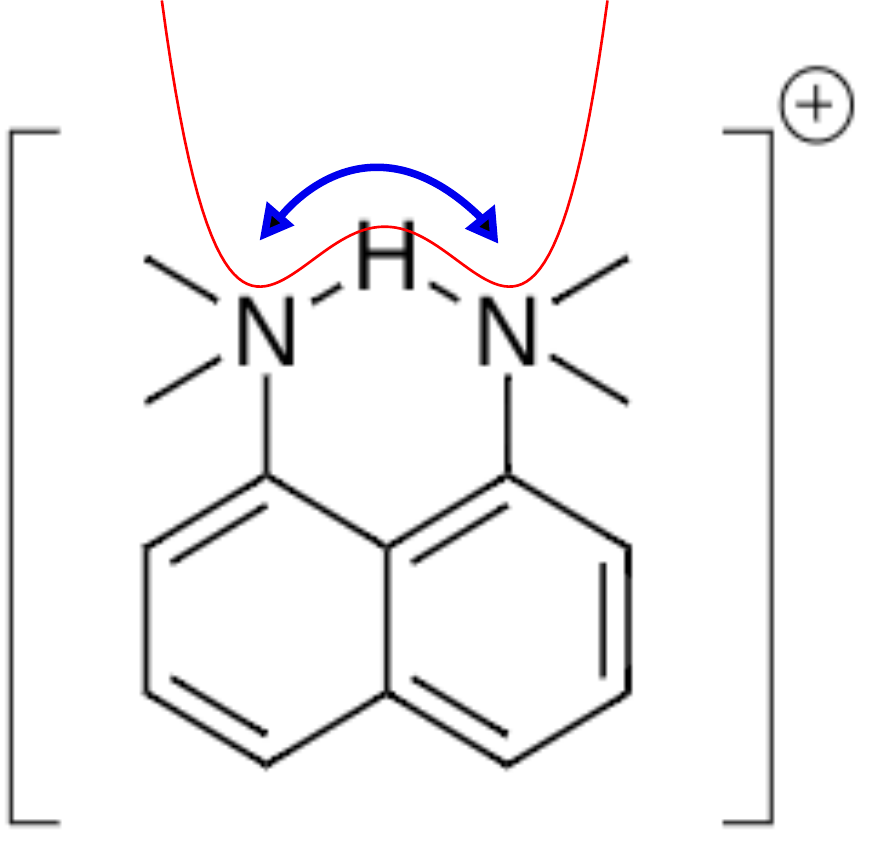}} 
\subfigure[]{\includegraphics[width=\columnwidth]{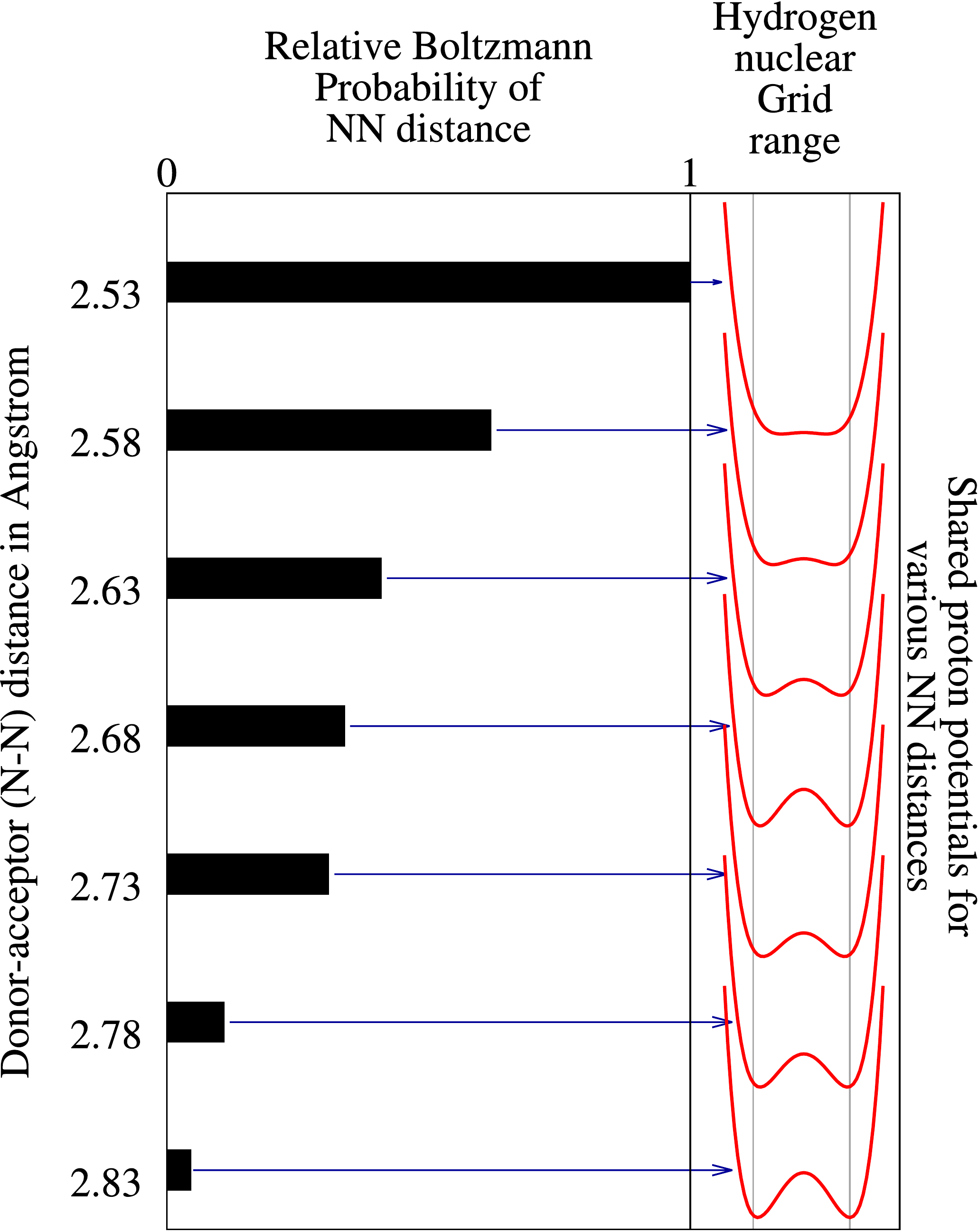}}
\caption{\label{H-transfer-example} Figure (a): The molecular geometry for DMANH$^+$ with the shared proton potential surface shown in red. The quantum mechanical nature of the shared proton allows it to be simultaneously present in both wells, and here we use Eq. (\ref{HIT-gen}) to simulate the behavior of this shared proton through out mapping protocol in Eq. (\ref{HMol-HIT-correspondence}). Figure (b): The change in double well potential (and barrier height) can be seen as a function of donor-acceptor (N--N) distance. The bar-heights show the classical Boltzmann population for each N--N distance.}
\end{figure}
\section{Performance of the mapping protocol for a symmetric hydrogen bonded system} 
\label{results}
We examine the 
map by simulating 
the quantum dynamics of the molecular system and the ion-trap dynamics, on classical hardware, independently. 
In 
doing so we study the time-evolution of the initial wavepacket states prepared in the respective permuted basis representations for the molecular and Ising model Hamiltonians. As stated, the parameters in the Ising Hamiltonian are determined, and thus controlled, by the pre-computed matrix elements of the molecular Hamiltonian. 
The specific intra-molecular proton transfer problem 
considered here is that in the protonated 1,8-bis(dimethylamino) naphthalene (DMANH$^+$) system shown in Figure \ref{H-transfer-example}(a). The DMAN molecule has an extremely large proton affinity of $242$ kcal/mol \cite{lias1984j}, with DMANH$^+$  $pK_{a}$ value in the range $12.1-12.3$ \cite{perrin1972dissociation}. As a result, the system is one of the most frequently investigated proton sponges. 
The NHN$^+$ hydrogen bond in proton sponges is attractive from the point of view of both the nature of the short potentially symmetric hydrogen-bond bridges\cite{SSHB-NielsonReview,SSHB-4,SSHB-Warshel-1,Sshb-Enzymes}, their infrared spectroscopic behavior and their propensity to occur in common nitrogen activation catalysts\cite{Schrock-N2-NCET,Hoffman-N2-H2-redelim}. Thus, the DMANH$^+$ system has been frequently studied as a model for short, low-barrier hydrogen bonds that have 
a role in certain enzyme-catalyzed reactions. In solution, the shared proton delocalization in DMANH$^+$ is controlled by a 
low-barrier symmetric double-well potential, with barrier height being influenced by solvent and temperature \cite{Jeremy-ref1,Jeremy-ref2}. 
In fact, the environment, variables such as solvent and temperature, influence the donor-acceptor distance fluctuations thus having a critical role on the quantum mechanical nature of the shared proton. The effect of these donor-acceptor variables is seen in Figure \ref{H-transfer-example}(b) where we present the shared-proton one-dimensional symmetric potentials (red-curves on right side of Figure \ref{H-transfer-example}(b)) for a range of donor-acceptor distances (left vertical axis in Figure \ref{H-transfer-example}(b)) with significant classical Boltzmann populations (black horizontally placed histograms in Figure \ref{H-transfer-example}(b)) at room temperature. Clearly, the barrier heights separating the minima in the red-curves as well as respective minimum energy positions are sensitive to donor-acceptor fluctuations and influence the spectroscopic properties of such hydrogen-bonded systems\cite{MeDH+,MeH+}. To emphasize this, in Figure \ref{H-transfer-example}(b), the light gray vertical lines are positioned to approximately coincide with the minimum energy values for the red potential energy surface at an NN distance of 2.83\AA\,. As the NN-distance gets smaller, the minimum energy points get closer to each other and changes the nature of the confinement potential in the shared hydrogen nucleus. Here, the effect of all of these aspects are studied by mapping the quantum nuclear dynamics problem on multiple potential surfaces, obtained from different donor-acceptor (NN) distances, to ion-trap quantum simulators.   

In the following subsections, we present the methods used to classically pre-compute the nuclear Hamiltonian for each of the donor-acceptor distances shown in Figure \ref{H-transfer-example}(b), and simulate the quantum nuclear dynamics on these potentials using the Ising model based ion-trap simulators.
We treat the shared proton stretch dimension within the 
Born-Oppenheimer limit. The nuclear Hamiltonian is determined by the ground electronic state 
potential energy surface. 
\subsection{Pre-computing the molecular Hamiltonian (${\cal{H}}^{Mol}$ in Eq. (\ref{HDAF-molQD})) on classical hardware}
In order to compute the potential energy surface for the intra-molecular proton transfer in the molecular system DMANH$^+$ (Figures \ref{Map-outline}a and \ref{H-transfer-example}(a)), we locate a symmetric stationary point with the shared proton at the center of the donor-acceptor axis. For the case of DMANH+, this stationary point turns out to be a transition state with one imaginary frequency that is obtained from the eigenstates of the electronic structure Hessian matrix, with the vibrational mode corresponding to the intra-molecular proton transfer direction. At this geometry, the shared proton is symmetrically located between the donor and acceptor nitrogen atoms. 
These calculations are performed using standard electronic structure methods. The level of electronic structure theory used is density functional theory with hybrid functional, B3LYP, and an atom-centered Gaussian basis set containing polarization and diffuse functions on all atoms, that is, 6-311++G(d,p). Future work will also include mapping of this Hamiltonian pre-computation step onto quantum hardware. A reduced dimensional potential energy surface calculation for one-dimensional proton motion along the donor-acceptor axis is performed at  the aforementioned stationary point geometry. This is also done for a set of donor acceptor distances with significant classical Boltzman populations as seen in Figure \ref{H-transfer-example}(b). The potential energy surfaces are obtained on a grid defined along the donor-acceptor axis. 
We choose $2^{N}$ number of equally spaced grid points, symmetrically located about the grid-center, and perform electronic structure calculations at these points, on a classical computing platform, at the level of theory mentioned above. The molecular Hamiltonian is computed (Eq. (\ref{HDAF-molQD})) and unitary transformed to achieve a block structure according to Section \ref{Transformations_Hmol}. 

\begin{figure}[htbp]
        \centering
       \includegraphics[width=\columnwidth]{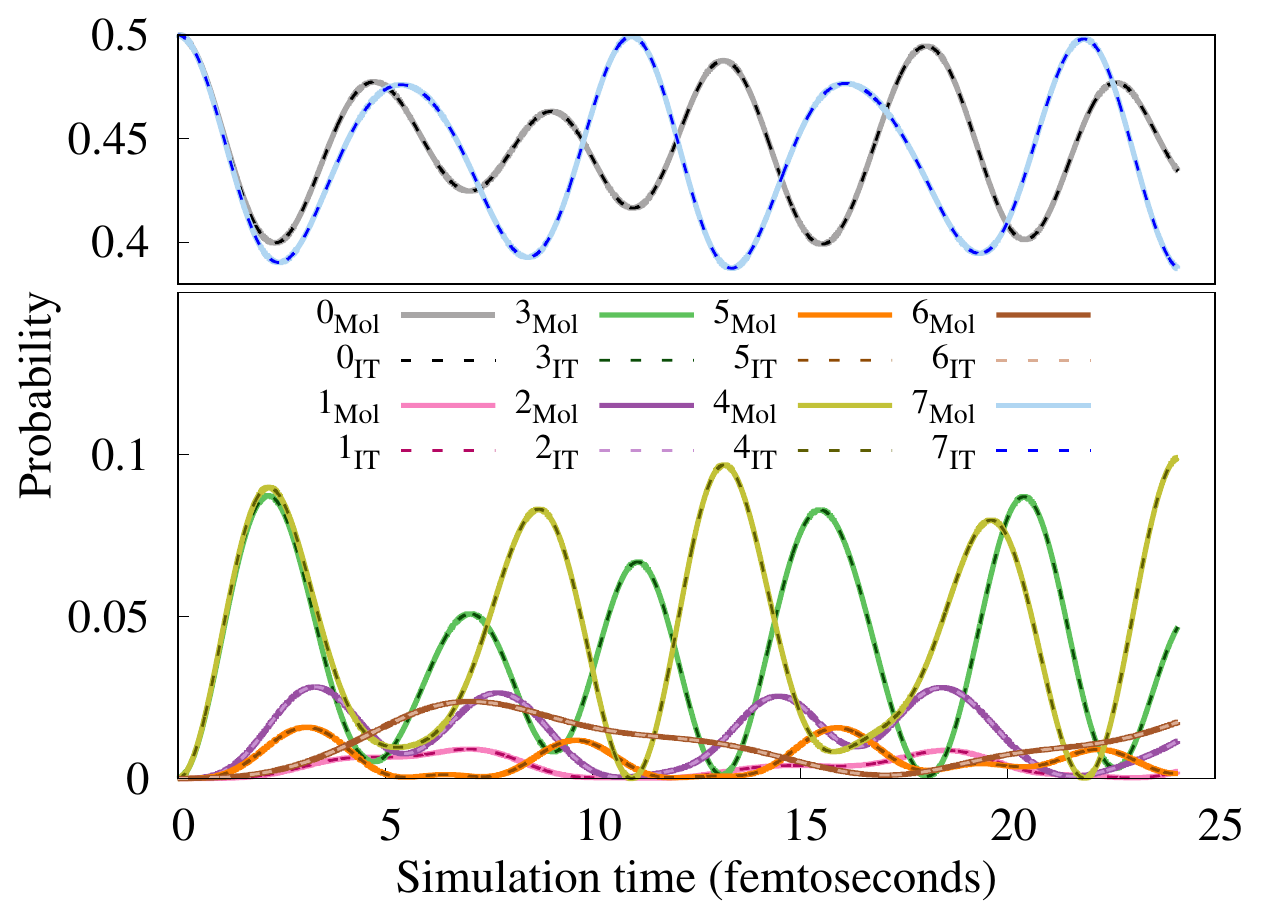}
        \caption{Dynamics of the molecular and the ion trap systems: The integer ($i$) depicts the projection of a  propagated state onto the $i^{th}$ permuted spin basis state and the corresponding Givens transformed grid basis state for the ion-trap (dashed) and the molecular system with $d_{DA} = 2.53$\AA\ (solid), respectively. Note that all propagation are conducted on classical platforms. 
        The agreement of the quantum dynamics in both systems is exact to within numerical round-off ($10^{-15}$). The two rows in the figure legend represent the two sets spanned by odd and even spin raising operators, $\left\{ {\bf S^+} \right\}$ acting on the $\ket{\downarrow \downarrow \cdots}$ spin state (dashed) and their corresponding Givens transformed grid basis states (solid) according to Eq. (\ref{lambda-x-correspondence}). An extended set of $d_{DA}$ are considered in Figure (\ref{fig:1}), and results for a longer term dynamics for the most stable structure ($d_{DA}= 2.53$\AA\ )is provided in Figure (\ref{fig:my_label}). 
        }
        \label{propagation_error}
\end{figure}

\begin{figure}[htbp]
    \subfigure[Projections of the time-dependent wavepackets onto Block I vectors]{\includegraphics[width=\columnwidth]{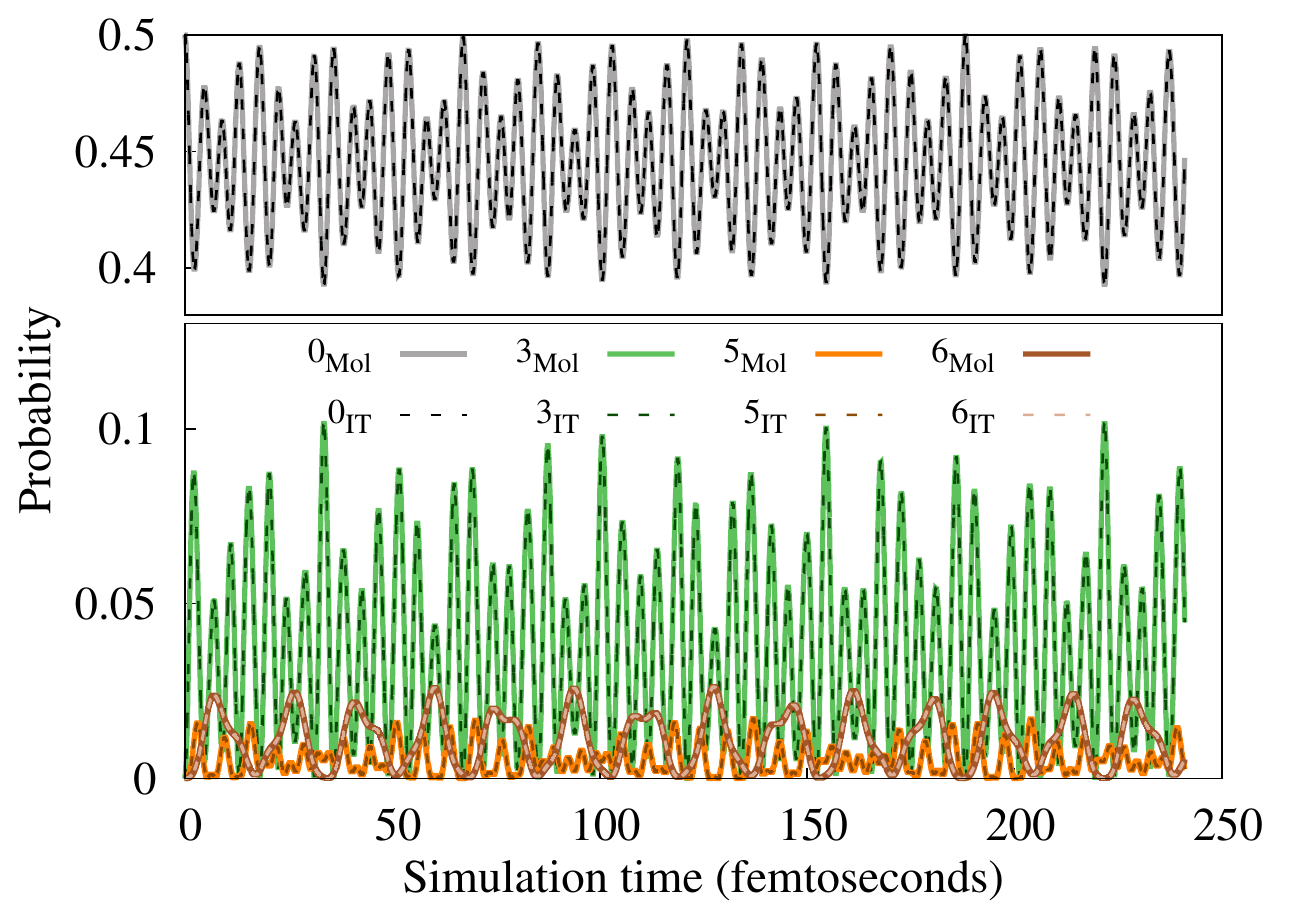}}
    \subfigure[Projections of the time-dependent wavepackets onto Block II vectors]{\includegraphics[width=\columnwidth]{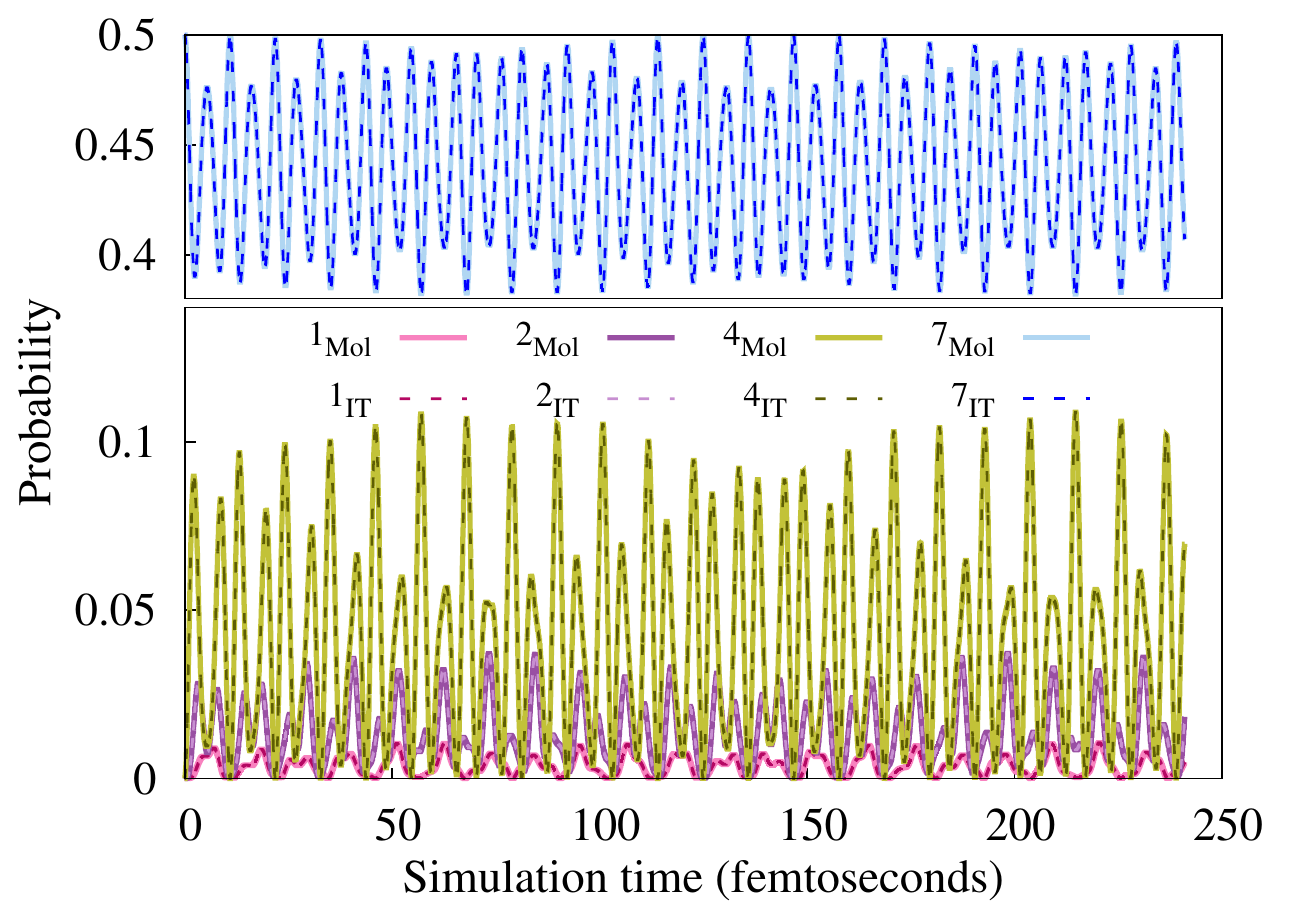}}
    \caption{The dynamics of the molecular system (solid) and the ion-trap system (dashed) that show their exact match to within numerical round-off ($10^{-15}$) over long simulation times sufficient to capture the molecular vibrational properties. Complements Figure \ref{propagation_error}.
    The projection of the respective time-dependent wavepackets onto basis vectors within each of the two decoupled blocks are shown separately for clarity.}
    \label{fig:my_label}
\end{figure}
   
\begin{figure*}[htbp]
\subfigure[$d_{DA} = 2.58$\AA\ \quad $\rho_{B} = 0.62$]{\includegraphics[width=7.25cm]{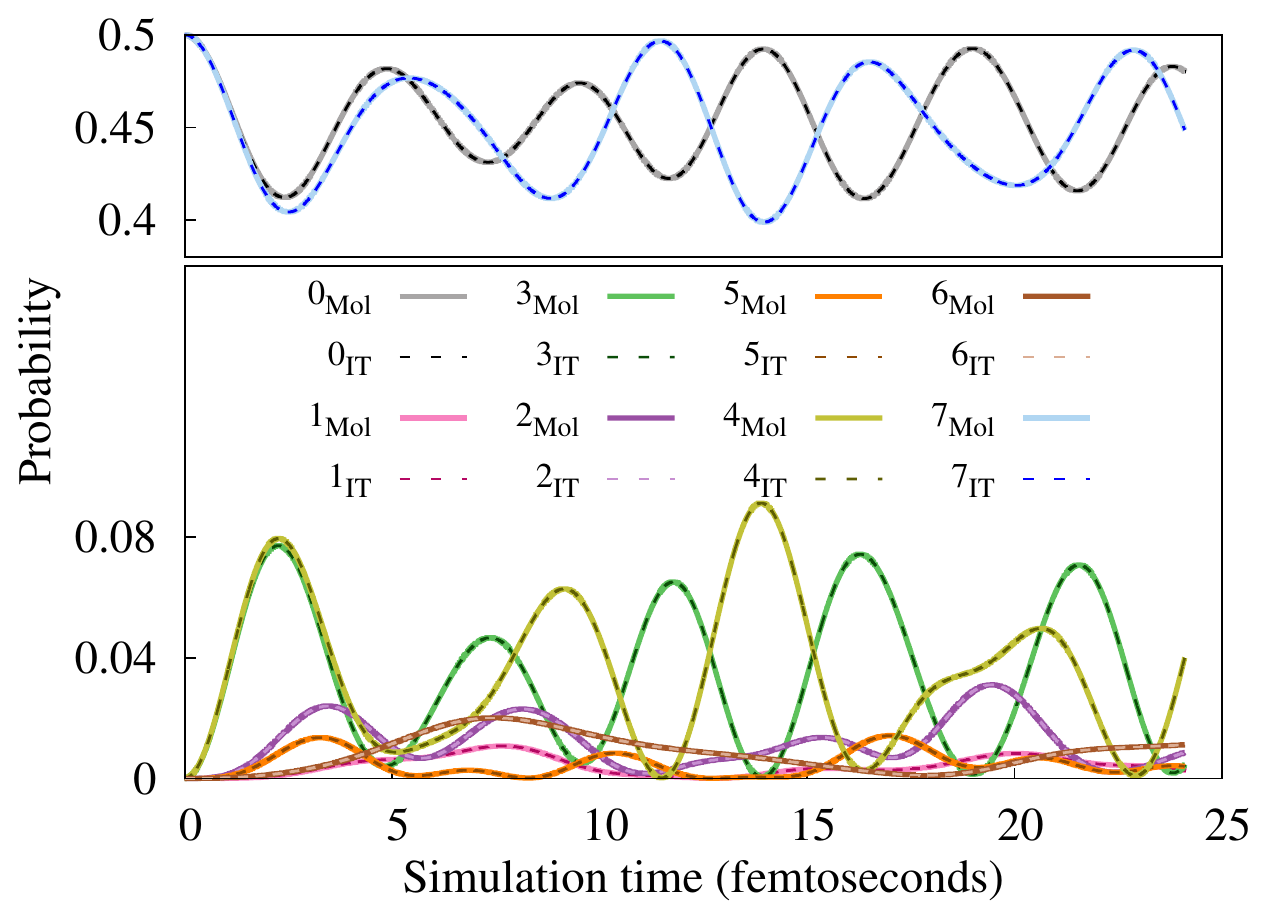}}
\subfigure[$d_{DA} = 2.63$\AA\ \quad $\rho_{B} = 0.41$]{\includegraphics[width=7.25cm]{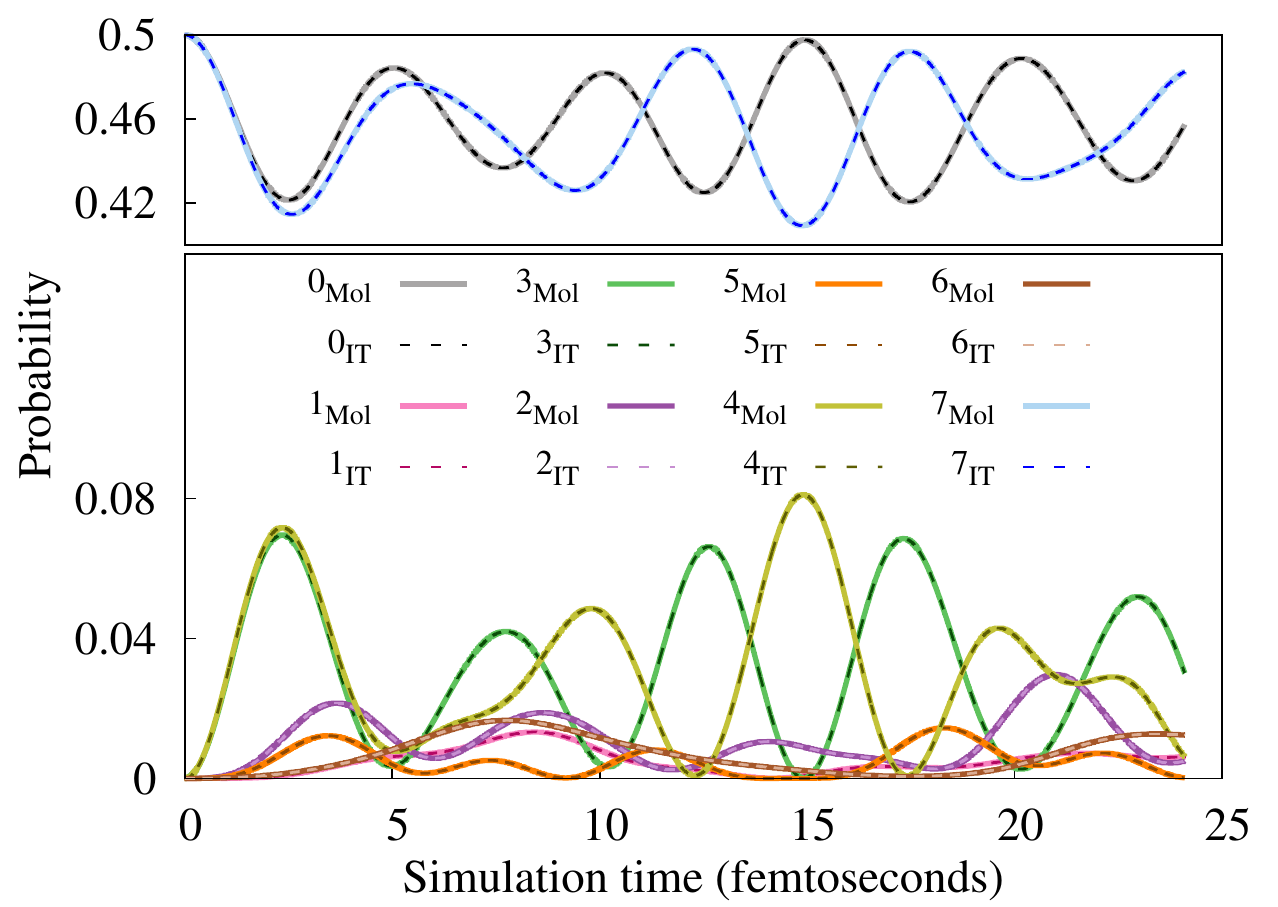}}
\subfigure[$d_{DA} = 2.68$\AA\ \quad $\rho_{B} = 0.34$]{\includegraphics[width=7.25cm]{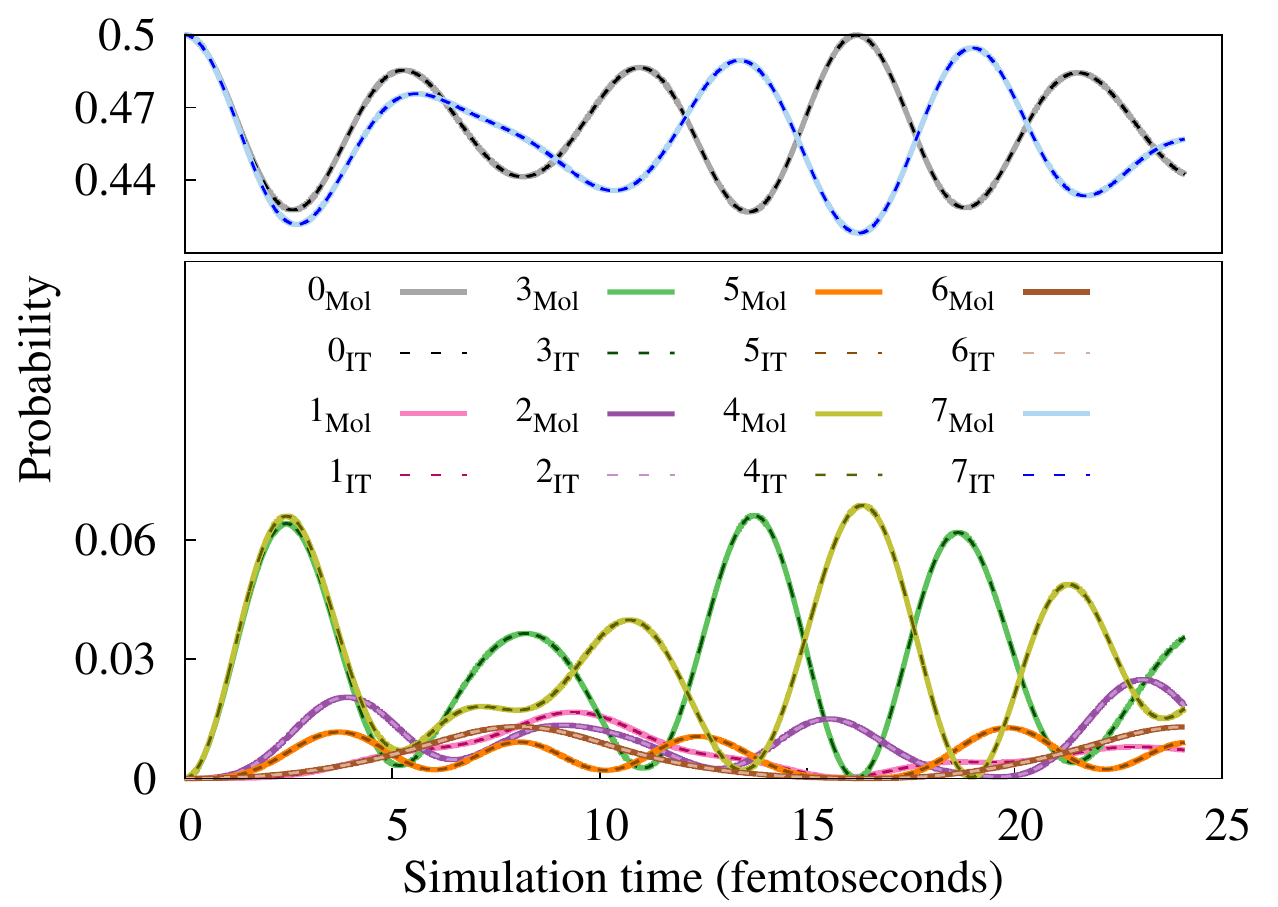}}
\subfigure[$d_{DA} = 2.73$\AA\ \quad $\rho_{B} = 0.39$]{\includegraphics[width=7.25cm]{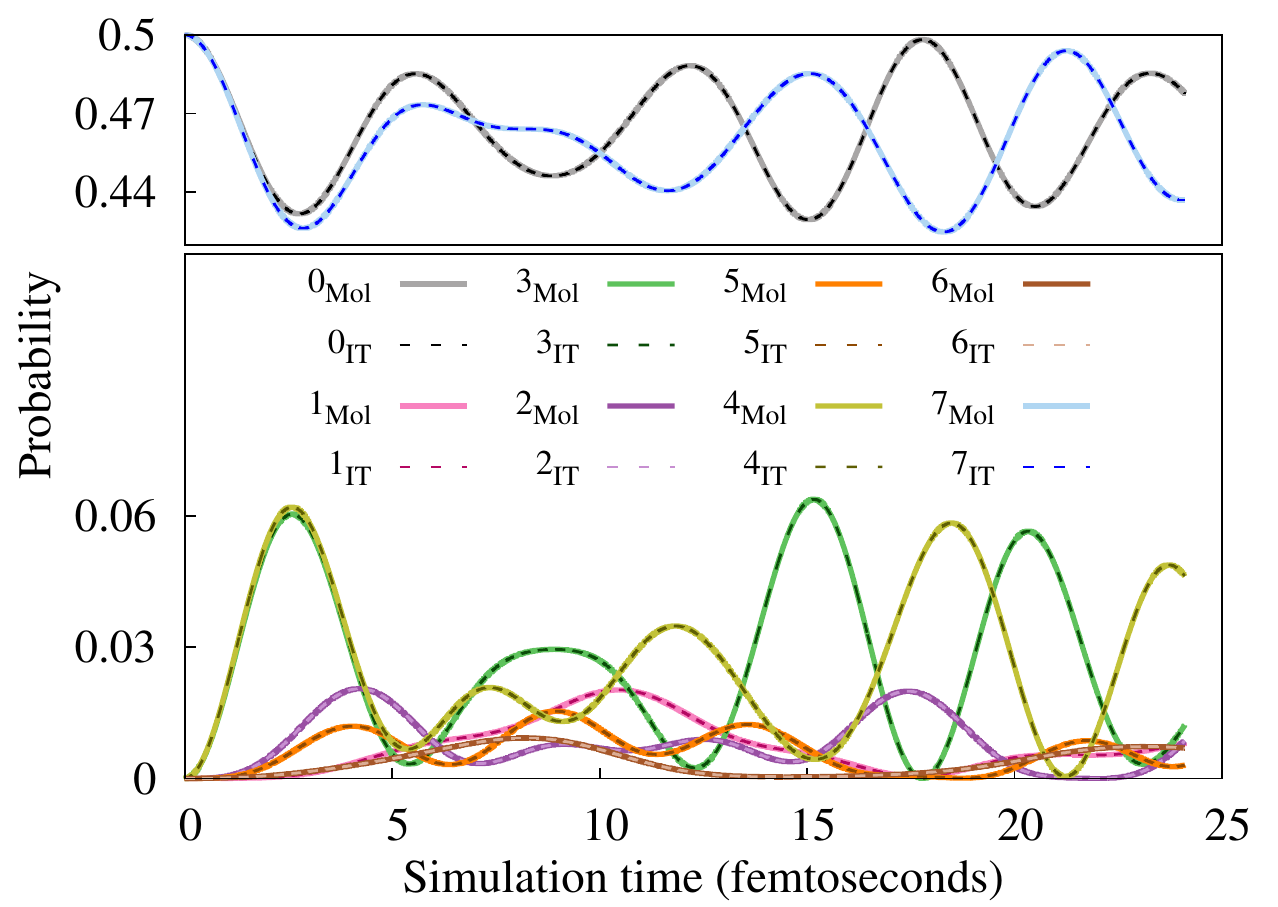}}
\subfigure[$d_{DA} = 2.78$\AA\ \quad $\rho_{B} = 0.11$]{\includegraphics[width=7.25cm]{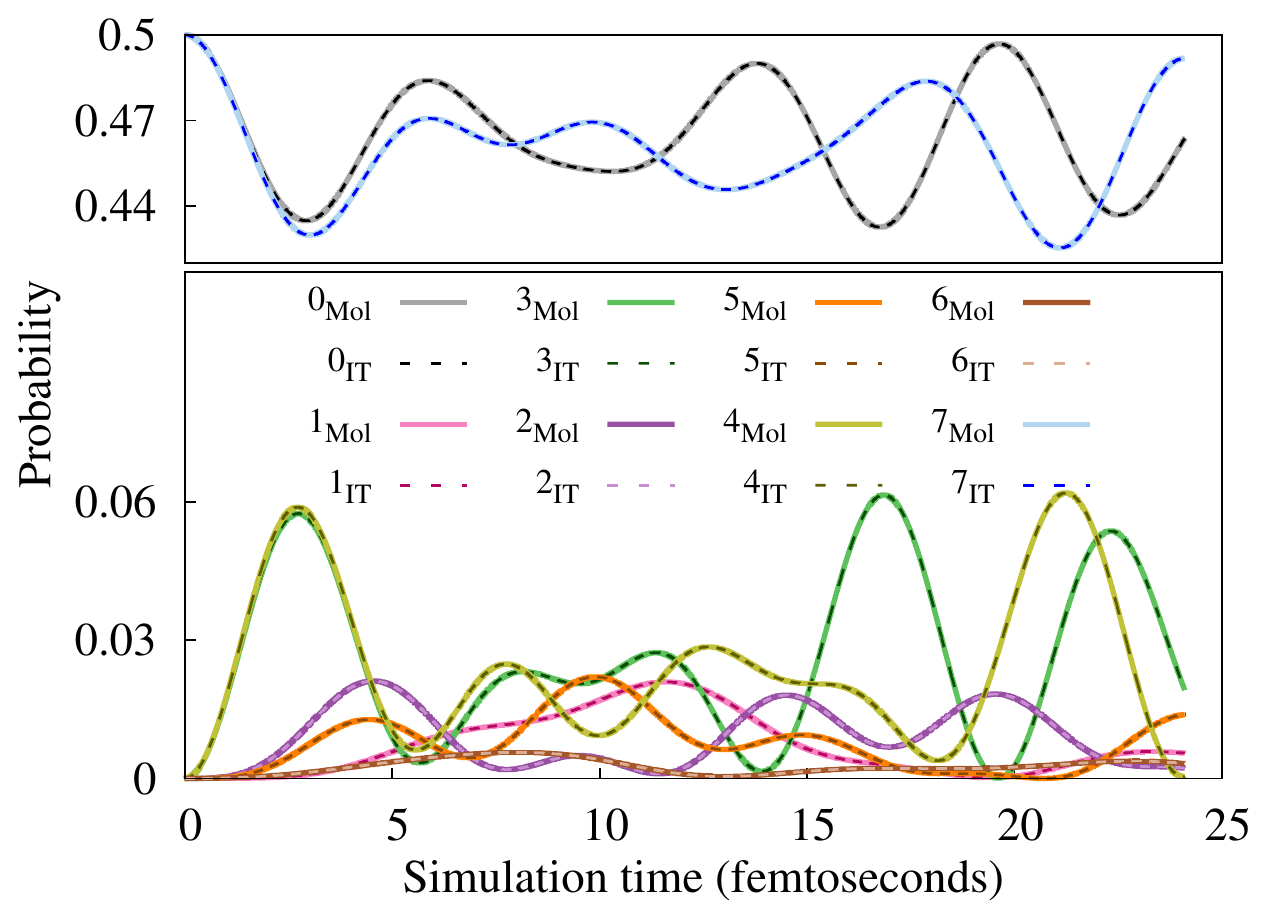}}
\subfigure[$d_{DA} = 2.83$\AA\ \quad $\rho_{B} = 0.047$]{\includegraphics[width=7.25cm]{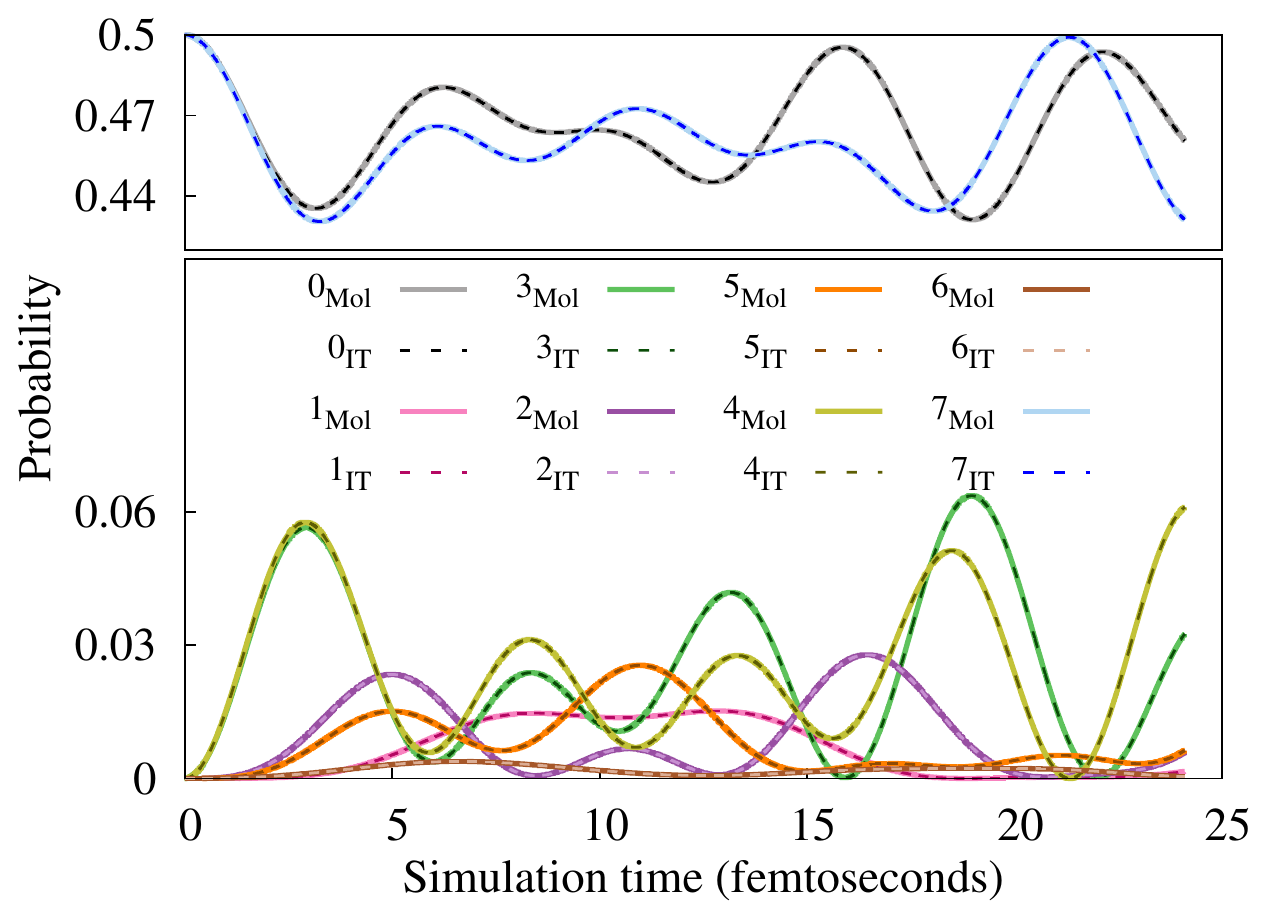}}
\caption{Similar to Figure \ref{propagation_error} 
but for multiple donor-acceptor distances ($d_{DA}$) between the nitrogen atoms for the molecule in Figure \ref{Map-outline}a. Boltzmann populations ($\rho_{B}$) are computed at 300K relative to the population of the configuration used in Figure \ref{propagation_error}, 
that has a $d_{DA}$ value of 2.53\AA\ . The correlated changes in the $\ket{0}$ and $\ket{7}$ projections, are clearly facilitated by components along other basis vectors, and these may have a critical role on the reactive process as a function of temperature. The fact that the ion-lattice dynamics displays the same dynamical trends provides an additional probe to complex chemical systems. } \label{fig:1}
\end{figure*}
\subsection{Quantum simulation of proton-transfer dynamics}
Given the block structure of both molecular and Ising Hamiltonians in the permuted and Givens transformed basis representations, 
the initial wavepacket for the ion-trap system is chosen as a coherent linear combination of the spin basis states: 
$\left\{\frac{\ket{\uparrow \uparrow \uparrow} + \ket{\downarrow \downarrow \downarrow}}{\sqrt{2}} \right\}$ on a three qubit system. 
Given the block structure of the Ising Hamiltonian 
with $\left\{ B_i^x, B_i^y \right\}$ turned off, the components of this initial state, $\ket{\uparrow \uparrow \uparrow} $ and $ \ket{\downarrow \downarrow \downarrow}$, are not coupled. 
Additionally, 
these states will not couple such as might be the case in the presence of $B_3^x-iB_3^y$ in the off-diagonal blocks: for example, pathways such as $\ket{\downarrow \downarrow \downarrow} \xrightarrow[]{B_3^x-iB_3^y} \ket{\uparrow \downarrow \downarrow} \xrightarrow[]{J_{12}^x-J_{12}^y} \ket{\uparrow \uparrow \uparrow}$ %
will remain unpopulated. Hence, in essence, $\ket{\uparrow \uparrow \uparrow}$ gets propagated as per the unitary evolution corresponding to the top diagonal block of the  Ising Hamiltonian 
and $ \ket{\downarrow \downarrow \downarrow}$ 
as per the bottom-block. 
This critical feature allows us to treat the two separated blocks as arising from two different ion-traps with two different sets of $\left\{ B_{i}^\gamma; J_{ij}^\gamma \right\}$ parameters.
Given the direct map in Eq. (\ref{lambda-x-correspondence}) between the permuted computational basis and the Givens transformed molecular grid basis, the initial wavepacket for the molecular system is to be chosen in an analogous fashion to the initial wavepacket of the ion-trap, which is $\left\{\frac{\ket{\tilde{x}_{0}} + \ket{\tilde{x}_{7}}}{\sqrt{2}} \right\}$. This essentially leads to the initial wavepacket for the quantum nuclear dynamics problem as being  chosen 
on one end of the grid, that is, a state localized closer to one of the nitrogen atoms in Figures \ref{Map-outline}a and \ref{H-transfer-example}a. This choice results in the initial nuclear wavepacket being symmetrically located at either end of the Givens transformed basis (Eq. (\ref{Givens-transformed-basis})).  
The spin-lattice and molecular wavepackets are then independently propagated for each potential obtained for different donor-acceptor separations, and compared to gauge accuracy of the quantum simulation. 

Given the recursive form of the matrix representation of the Ising Hamiltonian in Eq. (\ref{HIT-gen}), as discussed in Appendix \ref{Recursive_Structure_Ising} (see Eq. (\ref{HIT-nqubit-Block})), the ion-trap hardware initial wavepacket state is directly propagated by the choice of $\left\{ B_{i}^\gamma; J_{ij}^\gamma \right\}$ for arbitrary time-segments. In this publication we do not seek experimental validation using a real ion-trap simulator, but emulate the time-evolution of the ion-trap system according to the Hamiltonian in Eq. (\ref{HIT-gen}) on classical hardware, by using the eigenstates of the Ising Hamiltonian in Appendix \ref{Recursive_Structure_Ising}. The time-dependent probabilities resulting from the projection of the resultant time-dependent wavepacket on the computational basis, at each interval of time, is shown using dashed lines in Figures \ref{propagation_error} and \ref{fig:my_label} for a donor-acceptor distance value of $2.53$\AA\, and for the full set of donor-acceptor distance values in Figure \ref{fig:1}. (The donor-acceptor distance value of $2.53$\AA\, corresponds to the most stable structure, but as seen from Figure \ref{fig:1}, there are several other geometries that are also populated (at $300K$) even from a purely classical Boltzmann estimation.) 
Similarly, we determine the time-evolution of the initial wavepacket for the molecular system by using the eigenstates of the transformed Hamiltonian in Eq. (\ref{Htilde-il}), and the resulting probabilities from the projection of the time-dependent wavepacket on the Givens transformed grid basis, $\left\{ \ket{\tilde{x}} \right\}$ are shown using solid lines in figures \ref{propagation_error}, \ref{fig:my_label}, and \ref{fig:1}. 
The probabilities match exactly, apart from numerical round-off error ($10^{-15}$), for the quantum simulation of the dynamics of the two systems. Clearly, this is also true for much longer time intervals as can be seen in Figure \ref{fig:my_label}. 
Given the exact match between the spin-lattice dynamics and the quantum chemical dynamics, the 
features present in ion-trap dynamics must also exist in the chemical dynamics problem. Thus through the isomorphism constructed above, our algorithm allows the ability to probe any entanglement that may be present in chemical systems. 

\section{Conclusions and Outlook}
\label{concl}
The successful simulation of quantum nuclear dynamics on quantum hardware 
promises a new paradigm for 
studying a broader class of coupled electron nuclear transfer problems. 
In this publication, we provide a general, but approximate mapping procedure between a quantum chemical dynamics problem, constructed on a single Born-Oppenheimer surface, and an ion-trap quantum simulator where the dynamics is dictated by a generalized form of the Ising model Hamiltonian. The key step involved in facilitating our map is the partitioning of the coupled qubit space into two zones using only odd or even powers of the total spin raising operators that are used to generate such a coupled qubit space. Once the coupled qubit computational basis set is partitioned in such a way, the Ising model Hamiltonian reduces into a block form thus allowing the possibility to map {\em all} problems that may be written in a similar block form. In some sense, we have also taken here the necessary steps to detail the kinds of general problems that can be solved exactly on a quantum system whose dynamics is dictated by a generalized form of the Ising model Hamiltonian. In this particular paper, though, we consider a symmetric proton-transfer problem and then go on to show how such a problem can be mapped to an ion-trap system, and also show that the dynamics of the two systems is {\em identical} provided the parameters of the ion trap are chosen in concert with that of the molecular system obtained from classical pre-computation. We also provide error bounds for this approximate algorithm for arbitrary number of qubits. 

General quantum nuclear dynamics problems, however, have unsymmetric potential energy surfaces and are generally performed in higher dimensions. This work will become critical in extending our mapping protocol to general potentials in higher dimensions, as will be considered in future publications. 
In addition, the next set of steps also include inspection of nuclear wavepacket basis set dependence on the accuracy of the proposed map. Using appropriate basis sets, it may be possible to reduce the number of independent descriptors within the molecular Hamiltonian, thus tailoring the accuracy of the map according to the constraints provided in this paper. 

\section*{Acknowledgments} This research was supported by the National Science
Foundation grant OMA-1936353 to SSI, PR, JMS and AS. The authors are grateful to Dr. Miguel \'Angel \'Lopez Ruiz for valuable comments.

\appendix

\section{Recursive, block structure of the Ising Hamiltonian matrix}\label{Recursive_Structure_Ising} The Ising Hamiltonian matrix, $\mathbf{{H}_{N}}$, for a spin-lattice containing $N$ spin sites that represents Eq. (\ref{HIT-gen}), 
when represented in the aforementioned  computational basis set partitioned according to sets that independently span $ \left\{ {\bf S^+}^{2n}  \ket{2^{N}-1} \right\}$ and $\left\{ {\bf S^+}^{2n-1}  \ket{2^{N}-1} \right\}$, may be recursively written in a blocked form as,  
\begin{widetext}
\begin{align}
    \mathbf{{H}_{N}} 
    &= \begin{bmatrix}
    \mathbf{H_{N}^{D1}} & \mathbf{B_{N}} \\
    \mathbf{B_{N}^{\dagger}} & \mathbf{H_{N}^{D2}} \\
    \end{bmatrix} \nonumber \\ &= 
    \left[ \begin{array}{@{}c|c@{}} 
    \begin{matrix}
        \substack{\mathbf{{H}_{N-1}^{D1}} + B_{N}^{z}\mathbf{I_{2^{N-2}}} \\ +\mathbf{J_{z,N}^{1}}} & \mathbf{J_{xy,N}^{1}}\\
        \mathbf{J_{xy,N}^{1}}^{\top} & \substack{\mathbf{{H}_{N-1}^{D2}} - B_{N}^{z}\mathbf{I_{2^{N-2}}}  \\+ \mathbf{J_{z,N}^{1}}}
    \end{matrix}
    & \begin{matrix}
        \mathbf{B_{N-1}} & \left(B_{N}^{x} - \imath B_{N}^{y} \right)\mathbf{I_{2^{N-2}}}\\
        \left(B_{N}^{x} + \imath B_{N}^{y} \right)\mathbf{I_{2^{N-2}}} & \mathbf{B_{N-1}}
    \end{matrix}\\
    \hline
    \begin{matrix}
        \mathbf{B_{N-1}^{\dagger}} & \left(B_{N}^{x} + \imath B_{N}^{y} \right)\mathbf{I_{2^{N-2}}}\\
        \left(B_{N}^{x} - \imath B_{N}^{y} \right)\mathbf{I_{2^{N-2}}} & \mathbf{B_{N-1}^{\dagger}}
    \end{matrix}
    & \begin{matrix}
        \substack{\mathbf{{H}_{N-1}^{D2}} + B_{N}^{z}\mathbf{I_{2^{N-2}}} \\+ \mathbf{J_{z,N}^{2}}} & \mathbf{J_{N}^{2}}\\
        \mathbf{J_{xy,N}^{2}}^{\top} & \substack{\mathbf{{H}_{N-1}^{D1}} + B_{N}^{z}\mathbf{I_{2^{N-2}}}  \\+ \mathbf{J_{z,N}^{2}}}
    \end{matrix}
\end{array} \right]
\label{HIT-recursive}
\end{align}
\end{widetext}
where both the diagonal blocks $\mathbf{H_{N}^{D1}}$ and $\mathbf{H_{N}^{D2}}$, and the off-diagonal block $\mathbf{B_{N}}$ are recursively defined above, and  $\mathbf{I_{2^{N}}}$ denotes an identity matrix of size $2^{N}$. The $\mathbf{J_{xy,N}^{1}}$ and $\mathbf{J_{xy,N}^{2}}$ matrices that appear in the recursive definition of the diagonal blocks, labeled with superscripts $\bf{D_{1}}$ and $\bf{D_{2}}$ respectively, contain intersite coupling of the $N^{th}$ spin site with the remaining $N-1$ sites.
To arrive at the matrix elements belonging to  $\mathbf{J_{xy,N}^1}$ and $\mathbf{J_{xy,N}^2}$ above, 
a bitwise XOR operation is constructed between the corresponding computational bases, $\ket{\tilde{\lambda}}$  and $\ket{\tilde{\lambda}^\prime}$ coupled by the them. The XOR operation provides the identity of the spin sites where these computational basis vectors differ, that is when the spin states are flipped across $\ket{\tilde{\lambda}}$  and $\ket{\tilde{\lambda}^\prime}$.
When the bases differ at two spin lattice site locations, $i$ and $j$, the corresponding matrix element of $\mathbf{J_{xy,N}^1}$ or $\mathbf{J_{xy,N}^2}$ 
is given by $J_{ij}^{x}\pm J_{ij}^{y}$.
The phase preceding the corresponding  $J_{ij}^{y}$ values results from an XNOR operation on the $i, j$ lattice sites discovered through the XOR operation above. 

The terms, ${\mathbf{J_{z,N}^{1}}}$ and ${\mathbf{J_{z,N}^{2}}}$, in Eq. (\ref{HIT-recursive}) are also defined in a similar fashion. Both ${\mathbf{J_{z,N}^{1}}}$ and ${\mathbf{J_{z,N}^{2}}}$ matrices are diagonal in form. Thus, the diagonal elements of $\mathbf{H_{N}^{D1}}$ and $\mathbf{H_{N}^{D2}}$ are incremented by a linear combination of all possible intersite couplings of the $N^{th}$ spin site with the remaining $N-1$ sites given by, $\sum\limits_{i=1}^{N-1}(-1)^{\tilde{\lambda}_{i}\oplus \tilde{\lambda}_{N}}J_{iN}^{z}$. 

As noted in the main paper, setting all the  transverse local qubit magnetic fields, $B_i^x$ and $B_i^y$ to zero in Eq. (1), 
which may be recursively written as,
\begin{widetext}
\begin{align}
\mathbf{H_{N}} = 
\begin{bmatrix}
\substack{\mathbf{H_{N-1}^{D1}} + B_{N}^{z}\mathbf{I_{2^{N-2}}}\\+\mathbf{J_{z,N}^{1}}} & \mathbf{J_{xy,N}^{1}
} & \mathbf{0} & \mathbf{0}\\
\mathbf{J_{xy,N}^{1}
}^\top & \substack{\mathbf{H_{N-1}^{D2}} - B_{N}^{z}\mathbf{I_{2^{N-2}}}\\+\mathbf{J_{z,N}^{1}}} & \mathbf{0} & \mathbf{0}\\
\mathbf{0} & \mathbf{0} & \substack{\mathbf{H_{N-1}^{D2}} + B_{N}^{z}\mathbf{I_{2^{N-2}}}\\+\mathbf{J_{z,N}^{2}}} & \mathbf{J_{xy,N}^{2}
}\\
\mathbf{0} & \mathbf{0} & \mathbf{J_{xy,N}^{2}
}^\top & \substack{\mathbf{H_{N-1}^{D1}} - B_{N}^{z}\mathbf{I_{2^{N-2}}}\\+\mathbf{J_{z,N}^{2}}}\\
\end{bmatrix}
\label{HIT-nqubit-Block-1}
    \end{align}
\end{widetext}
    
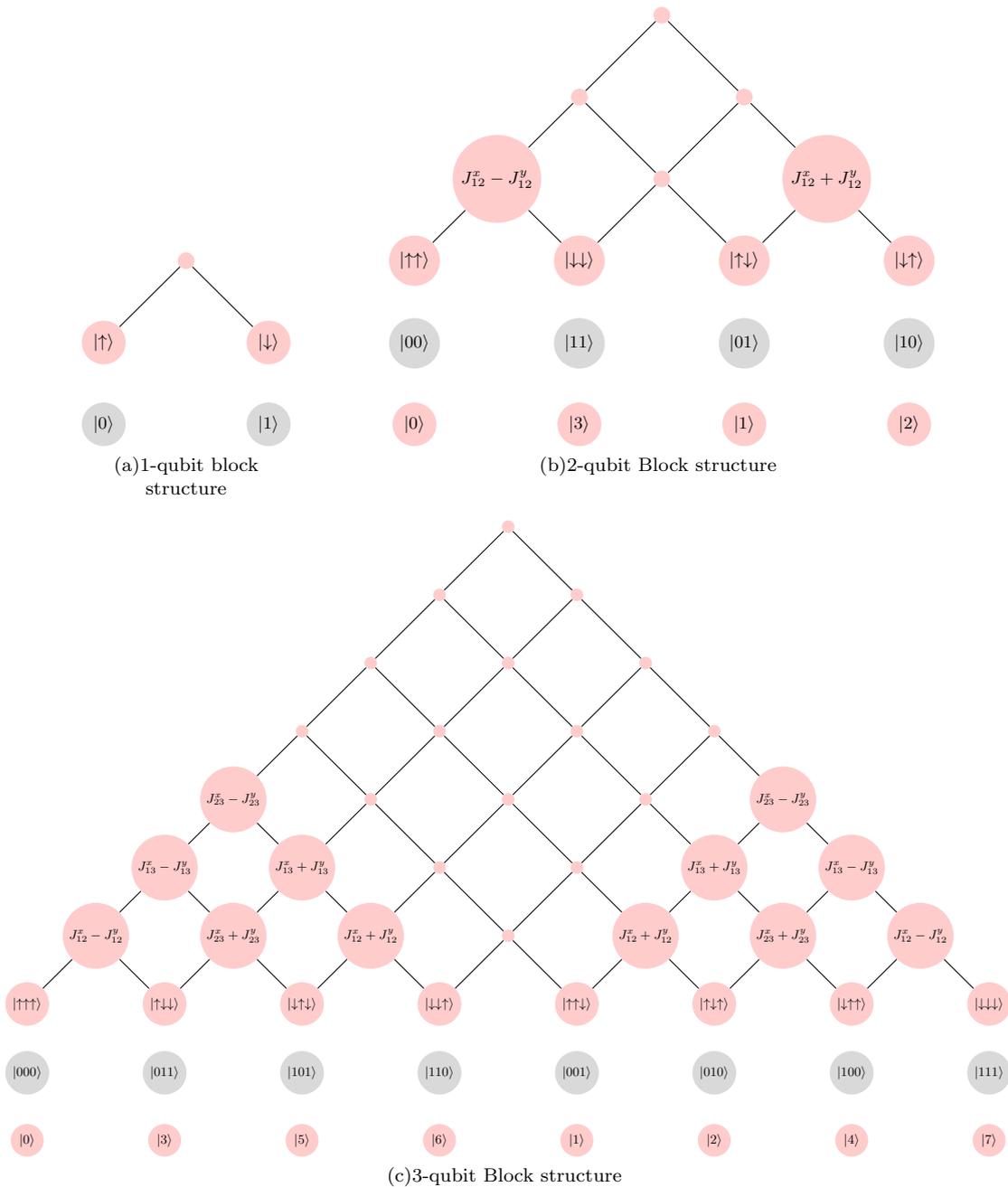
\begin{figure*}[htb!]
\subfigure[1-qubit block structure]{
\begin{tikzpicture}[scale=1.2,auto=center,every node/.style={circle,scale=0.8,fill=red!20}]
    %
    \node [fill=gray!30](a1) at (1,0) {$\ket{0}$};  
    \node [fill=gray!30](a2) at (3,0) {$\ket{1}$};
    \node (a1) at (1,1) {$\ket{\uparrow}$};  
    \node (a2) at (3,1) {$\ket{\downarrow}$};
    \node (a5) at (2,2) {};
    \draw (a1) -- (a5);
    \draw (a2) -- (a5);
\end{tikzpicture}
} \hspace{1cm}
\subfigure[2-qubit Block structure]{
\begin{tikzpicture}
[scale=1.2,auto=center,every node/.style={circle,scale=0.8,fill=red!20}] 
    \node [fill=gray!30](a1) at (1,-0.01) {$\ket{00}$};  
    \node [fill=gray!30](a2) at (3,-0.01) {$\ket{11}$};
    \node [fill=gray!30](a3) at (5,-0.01) {$\ket{01}$};
    \node [fill=gray!30](a4) at (7,-0.01) {$\ket{10}$};
    \node (a1) at (1,-1) {$\ket{0}$};  
    \node (a2) at (3,-1) {$\ket{3}$};
    \node (a3) at (5,-1) {$\ket{1}$};
    \node (a4) at (7,-1) {$\ket{2}$};
    \node (a1) at (1,1) {$\ket{\uparrow \uparrow}$};  
    \node (a2) at (3,1) {$\ket{\downarrow \downarrow}$};
    \node (a3) at (5,1) {$\ket{\uparrow \downarrow}$};
    \node (a4) at (7,1) {$\ket{\downarrow \uparrow}$};
    \node (a5) at (2,2) {$J_{12}^{x}-J_{12}^{y}$};
    \node (a6) at (6,2) {$J_{12}^{x}+J_{12}^{y}$};
    \node (a13) at (4,2) { };
    \node (a15) at (3,3) { };
    \node (a16) at (5,3) { };
    \node (a17) at (4,4) { };
    \draw (a1) -- (a5);
    \draw (a5) -- (a15);
    \draw (a15) -- (a17);
    \draw (a2) -- (a5);
    \draw (a2) -- (a13);
    \draw (a3) -- (a13);
    \draw (a3) -- (a6);
    \draw (a6) -- (a4);
    \draw (a6) -- (a16);
    \draw (a13) -- (a15);
    \draw (a13) -- (a16);
    \draw (a16) -- (a17);
\end{tikzpicture}}
\subfigure[3-qubit Block structure]{
\begin{tikzpicture}
[scale=1,auto=center,every node/.style={circle,scale=0.6,fill=red!20}]
    \node [fill=gray!30](l1) at (1,-0.01) {$\ket{000}$};  
    \node [fill=gray!30](l2) at (3,-0.01) {$\ket{011}$};        
    \node [fill=gray!30](l3) at (5,-0.01) {$\ket{101}$};
    \node [fill=gray!30](l4) at (7,-0.01) {$\ket{110}$};
    \node [fill=gray!30](l7) at (9,-0.01) {$\ket{001}$};  
    \node [fill=gray!30](l8) at (11,-0.01) {$\ket{010}$};        
    \node [fill=gray!30](l9) at (13,-0.01) {$\ket{100}$};
    \node [fill=gray!30](l10) at (15,-0.01) {$\ket{111}$};
    \node (a1) at (1,1) {$\ket{\uparrow \uparrow \uparrow}$};  
    \node (a2) at (3,1) {$\ket{\uparrow \downarrow \downarrow}$};    
    \node (a3) at (5,1) {$\ket{\downarrow \uparrow \downarrow}$};
    \node (a4) at (7,1) {$\ket{\downarrow \downarrow \uparrow}$};
    \node (a7) at (9,1) {$\ket{\uparrow \uparrow \downarrow}$};  
    \node (a8) at (11,1) {$\ket{\uparrow\downarrow\uparrow}$};   
    \node (a9) at (13,1) {$\ket{\downarrow\uparrow\uparrow}$};
    \node (a10) at (15,1) {$\ket{\downarrow\downarrow\downarrow}$};
    \node (a36) at (1,-1) {$\ket{0}$};  
    \node (a37) at (3,-1) {$\ket{3}$};   
    \node (a38) at (5,-1) {$\ket{5}$};
    \node (a39) at (7,-1) {$\ket{6}$};
    \node (a40) at (9,-1) {$\ket{1}$};  
    \node (a41) at (11,-1) {$\ket{2}$};  
    \node (a42) at (13,-1) {$\ket{4}$};
    \node (a43) at (15,-1) {$\ket{7}$};    \node  (a5) at (2,2) {$J_{12}^{x}-J_{12}^{y}$};
    \node  (a6) at (6,2) {$J_{12}^{x}+J_{12}^{y}$};
    \node  (a13) at (4,2) {$J_{23}^{x}+J_{23}^{y}$};
    \node (a15) at (3,3) {$J_{13}^{x}-J_{13}^{y}$};
    \node (a16) at (5,3) {$J_{13}^{x}+J_{13}^{y}$};
    \node (a17) at (4,4) {$J_{23}^{x}-J_{23}^{y}$};
    \draw (a1) -- (a5);
    \draw (a5) -- (a15);
    \draw (a15) -- (a17);
    \draw (a2) -- (a5);
    \draw (a2) -- (a13);
    \draw (a3) -- (a13);
    \draw (a3) -- (a6);
    \draw (a6) -- (a4);
    \draw (a6) -- (a16);
    \draw (a13) -- (a15);
    \draw (a13) -- (a16);
    \draw (a16) -- (a17);
%
    \node (a11) at (10,2) {$J_{12}^{x}+J_{12}^{y}$};
    \node (a12) at (14,2) {$J_{12}^{x}-J_{12}^{y}$};
    \node (a14) at (12,2) {$J_{23}^{x}+J_{23}^{y}$};
    \node (a18) at (12,4) {$J_{23}^{x}-J_{23}^{y}$};
    \node (a19) at (11,3) {$J_{13}^{x}+J_{13}^{y}$};
    \node (a20) at (13,3) {$J_{13}^{x}-J_{13}^{y}$};
    \draw (a7) -- (a11);
    \draw (a8) -- (a11);
    \draw (a9) -- (a12);
    \draw (a10) -- (a12);
    \draw (a8) -- (a14);
    \draw (a9) -- (a14);
    \draw (a14) -- (a19);
    \draw (a14) -- (a20);
    \draw (a11) -- (a19);
    \draw (a12) -- (a20);
    \draw (a20) -- (a18);
    \draw (a19) -- (a18);
    \node (a21) at (8,2) {    };
    \node (a22) at (7,3) {};
    \node (a23) at (9,3) {};
    \node (a24) at (6,4) {};
    \node (a25) at (8,4) {  };
    \node (a26) at (10,4) {};
    \node (a27) at (5,5) {};
    \node (a28) at (7,5) {};
    \node (a29) at (9,5) {};
    \node (a30) at (11,5) {};
    \node (a31) at (6,6) {};
    \node (a32) at (8,6) {   };
    \node (a33) at (10,6) {};
    \node (a34) at (8,8) {    };
    \node (a35) at (7,7) {};
    \node (a36) at (9,7) {};
    \draw (a17) -- (a27);
    \draw (a16) -- (a24);
    \draw (a6) -- (a22);
    \draw (a4) -- (a21);
    \draw (a31) -- (a27);
    \draw (a28) -- (a24);
    \draw (a25) -- (a22);
    \draw (a23) -- (a21);
    \draw (a31) -- (a35);
    \draw (a28) -- (a32);
    \draw (a25) -- (a29);
    \draw (a23) -- (a26);
    \draw (a34) -- (a35);
    \draw (a36) -- (a32);
    \draw (a33) -- (a29);
    \draw (a30) -- (a26);
    \draw (a7) -- (a21);
    \draw (a11) -- (a23);
    \draw (a19) -- (a26);
    \draw (a18) -- (a30);
    \draw (a22) -- (a21);
    \draw (a25) -- (a23);
    \draw (a29) -- (a26);
    \draw (a33) -- (a30);
    \draw (a22) -- (a24);
    \draw (a25) -- (a28);
    \draw (a29) -- (a32);
    \draw (a33) -- (a36);
    \draw (a34) -- (a36);
    \draw (a27) -- (a24);
    \draw (a31) -- (a28);
    \draw (a35) -- (a32);
\end{tikzpicture}}
\caption{\label{3qubit-block-graph} Complements Figure \ref{3qubit-block-graph-main}. 
At the base of each figure are the computational basis state kets. 
The interaction between any two states, $\ket{\tilde{\lambda}_i}$ and $\ket{\tilde{\lambda}_j}$  
can be read off from the graph, by starting at the two states and following the lines to their intersection. The node at the intersection 
gives the interaction between the two. For example, $\ket{2}$ and $\ket{7}$ in Figure (c) have an off-diagonal matrix element of $\left[ J_{13}^{x} - J_{13}^{y} \right]$. The blank nodes are zero, and show the block diagonal form of the Ising Hamiltonian when $\left\{ B_{i}^x; B_{i}^y \right\}$ are set to zero. 
}
\end{figure*}

At this stage it is critical to realize that the two blocks in the equation above are completely decoupled and basis vector components that undergo unitary evolution due to the top block are never influenced by elements from the bottom block and vice versa. This is explicitly elaborated in Figure \ref{3qubit-block-graph} for the case of two and three qubits. This presents us with an additional degree of flexibility for our quantum simulation. We exercise this  flexibility here 
and map separately the top and bottom blocks of the  equation above, to two different $N$-qubit ion trap systems controlled by parameters $\left\{ B^{z}_{i}; J^{\gamma}_{ij} \right\}$ and 
$\left\{ \tilde{B}^{z}_{i}; \tilde{J}^{\gamma}_{ij} \right\}$ respectively. It is important to note here, that while the underlying structure of each block in the Ising Hamiltonian matrix remains the same, two different sets of ion-trap control parameters are used to simulate the top and bottom blocks respectively, thus providing greater flexibility in simulating real systems. We, therefore, introduce a subtle change in  denoting the corresponding Ising model Hamiltonian as $\mathbf{{\cal H}_{N}}$, and allow the diagonal blocks to be independently determined in the following manner:
\begin{widetext}
\begin{align}
\mathbf{{\cal H}_{N}}
    &= \begin{bmatrix}
    \mathbf{H_{N}^{D1}} & \mathbf{0} \\
    \mathbf{0} & \mathbf{\tilde{H}_{N}^{D2}} \\
    \end{bmatrix} \nonumber \\ &= 
\begin{bmatrix}
\substack{\mathbf{H_{N-1}^{D1}} + B_{N}^{z}\mathbf{I_{2^{N-2}}}\\+\mathbf{J_{z,N}^{1}}} & \mathbf{J_{xy,N}^{1}
} & \mathbf{0} & \mathbf{0}\\
{\mathbf{J_{xy,N}^{1^{\top}}
}} & \substack{\mathbf{H_{N-1}^{D2}} - B_{N}^{z}\mathbf{I_{2^{N-2}}}\\+\mathbf{J_{z,N}^{1}}} & \mathbf{0} & \mathbf{0}\\
\mathbf{0} & \mathbf{0} & \substack{\mathbf{\Tilde{H}_{N-1}^{D2}} + \Tilde{B}_{N}^{z}\mathbf{I_{2^{N-2}}}\\+\mathbf{\tilde{J}_{z,N}^{2}}} & \mathbf{\Tilde{J}_{xy,N}^{2}
}\\
\mathbf{0} & \mathbf{0} & \mathbf{\Tilde{J}_{xy,N}^{2^{\top}}
} & \substack{\mathbf{\Tilde{H}_{N-1}^{D1}} - \Tilde{B}_{N}^{z}\mathbf{I_{2^{N-2}}}\\+\mathbf{\tilde{J}_{z,N}^{2}}}\\
\end{bmatrix}
\label{HIT-nqubit-Block}
    \end{align}
\end{widetext}
    \noindent where the top-block is controlled by parameters, $\left\{ B^{z}_{i}; J^{\gamma}_{ij} \right\}$, whereas the bottom block is controlled by a different set of ion-trap parameters, $\left\{ \tilde{B}^{z}_{i}; \tilde{J}^{\gamma}_{ij} \right\}$. The molecular Hamiltonian is  mapped to the above form of the Ising model Hamiltonian matrix.

We now illustrate the above form of Ising Hamiltonian for the 2- and 3-qubit systems. But, we note that the aforementioned basis set partitioning and Hamiltonian structure is completely general and applies to all cases. Explicitly written, for the case of two-qubits, Eq. (\ref{HIT-nqubit-Block}) 
takes the form
\begin{widetext}
\begin{equation}
    {\cal H}_{2} = \begin{bmatrix}
    B^{z}_{1} + B^{z}_{2} + J^{z}_{12} & J^{x}_{12} - J^{y}_{12} & 0 & 0\\
    J^{x}_{12} - J^{y}_{12} &  -B^{z}_{1} - B^{z}_{2} + J^{z}_{12}& 0 & 0\\
    0 & 0 &  \tilde{B}^{z}_{2} - \tilde{B}^{z}_{1} - \tilde{J}^{z}_{12}& \tilde{J}^{x}_{12} + \tilde{J}^y_{12}\\
    0 & 0 & \tilde{J}^{x}_{12} + \tilde{J}^{y}_{12} & \tilde{B}^{z}_{1} - \tilde{B}^{z}_{2} - \tilde{J}^{z}_{12}\\
    \end{bmatrix}
    \label{2qubit_BD_IT}
\end{equation}
\end{widetext}
where again we have highlighted the distinction between ion-trap simulators that represent the top block, $\left\{ B^{z}_{i}; J^{\gamma}_{ij} \right\}$, and those that control the bottom block, $\left\{ \tilde{B}^{z}_{i}; \tilde{J}^{\gamma}_{ij} \right\}$.
The three-qubit Hamiltonian is then recursively obtained from the two-qubit Hamiltonian as prescribed by Eq. (\ref{HIT-nqubit-Block}) and may be written in compact form as follows:
\begin{equation}
    {\cal H}_{3} = \begin{bmatrix}
    \substack{\mathbf{H_{2}^{D1}} + B_{3}^{z}\mathbf{I_{2}}\\+\mathbf{J_{z,3}^{1}}} & \mathbf{J_{xy,3}^{1}} & \mathbf{0} & \mathbf{0}\\
\mathbf{J_{xy,3}^{1^{\top}}} & \substack{\mathbf{H_{2}^{D2}} - B_{3}^{z}\mathbf{I_{2}}\\+\mathbf{J_{z,3}^{1}}} & \mathbf{0} & \mathbf{0}\\
\mathbf{0} & \mathbf{0} & \substack{\mathbf{\tilde{H}_{2}^{D2}} + \tilde{B}_{3}^{z}\mathbf{I_{2}}\\+\mathbf{\tilde{J}_{z,3}^{2}}} & \mathbf{\tilde{J}_{xy,3}^{2}}\\
\mathbf{0} & \mathbf{0} & \mathbf{\tilde{J}_{xy,3}^{2^{\top}}} & \substack{\mathbf{\tilde{H}_{2}^{D1}} - \tilde{B}_{3}^{z}\mathbf{I_{2}}\\+\mathbf{\tilde{J}_{z,3}^{2}}}\\
\end{bmatrix}
\end{equation}
Here, $\mathbf{H_{2}^{D1}}$ and $\mathbf{H_{2}^{D2}}$ refer to the top and bottom diagonal blocks of the two-qubit Ising Hamiltonian (Eq. (\ref{2qubit_BD_IT})) simulated using the ion-trap parameters $\left\{ B^{z}_{i}; J^{\gamma}_{ij} \right\}$ while $\mathbf{\tilde{H}_{2}^{D1}}$ and $\mathbf{\tilde{H}_{2}^{D2}}$ refer to the top and bottom blocks of a the two-qubit Ising Hamiltonian (Eq.\ref{2qubit_BD_IT}) controlled by $\left\{ \tilde{B}^{z}_{i}; \tilde{J}^{\gamma}_{ij} \right\}$. While most of $\mathbf{H_{2}^{D1}}$ and $\mathbf{H_{2}^{D2}}$ is preserved and appear as diagonal blocks of the 2 qubit Hamiltonian, the $N^{th}$ qubit on-site term $B_{3}^{z}$, and intersite coupling terms with all $N-1$ qubits $J^{z}_{13}$, $J^{z}_{23}$ with appropriate phases are added to each diagonal element. The quantities, $\mathbf{J_{3}^{1}}$, $\mathbf{J_{3}^{2}}$ in the top block and $\mathbf{\tilde{J}_{3}^{1}}$, $\mathbf{\tilde{J}_{3}^{2}}$ in the bottom block 
capture the interaction of qubits 1 and 2 with qubit 3 in the form of the inter-site coupling terms for the two ion-traps respectively. Explicitly,   
the 3 qubit system Hamiltonian becomes:
\begin{subequations}
\begin{align}
    {\cal H}_{3} = \mathbf{H_{3}^{D1}} \oplus \mathbf{\tilde{H}_{3}^{D2}}
\end{align}
where, for compactness we have written the ion-trap Hamiltonian as a direct sum of
\begin{widetext}
\begin{align}
    \mathbf{H_{3}^{D1}} = \begin{bmatrix}
    \substack{B^{z}_{1} + B^{z}_{2} + B^{z}_{3} \\ + J^{z}_{12} + J^{z}_{13} + J^{z}_{23}} &   J^{x}_{12} - J^{y}_{12}&   J^{x}_{13} - J^{y}_{13}&   J^{x}_{23} - J^{y}_{23}\\
J^{x}_{12} - J^{y}_{12}& \substack{B^{z}_{3} - B^{z}_{2} - B^{z}_{1} \\+J^{z}_{12} - J^{z}_{13} - J^{z}_{23}}&   J^{x}_{23} + J^{y}_{23}&   J^{x}_{13} + J^{y}_{13}\\
J^{x}_{13} - J^{y}_{13}&   J^{x}_{23} + J^{y}_{23}& \substack{B^{z}_{2} - B^{z}_{1} - B^{z}_{3} \\-J^{z}_{12} + J^{z}_{13} - J^{z}_{23}}&   J^{x}_{12} + J^{y}_{12}\\
J^{x}_{23} - J^{y}_{23}&   J^{x}_{13} + J^{y}_{13}&   J^{x}_{12} + J^{y}_{12}& \substack{B^{z}_{1} - B^{z}_{2} - B^{z}_{3} \\-J^{z}_{12} - J^{z}_{13} + J^{z}_{23}}\\
    \end{bmatrix}
\end{align}
\begin{align}
    \mathbf{\tilde{H}_{3}^{D2}} = \begin{bmatrix}
   \substack{\tilde{B}^{z}_{2} - \tilde{B}^{z}_{1} + \tilde{B}^{z}_{3} \\-\tilde{J}^{z}_{12} - \tilde{J}^{z}_{13} + \tilde{J}^{z}_{23}}&   \tilde{J}^{x}_{12} + \tilde{J}^{y}_{12}&   \tilde{J}^{x}_{13} + \tilde{J}^{y}_{13}&     \tilde{J}^{x}_{23} - \tilde{J}^{y}_{23}\\
   \tilde{J}^{x}_{12} + \tilde{J}^{y}_{12}&\substack{\tilde{B}^{z}_{1} - \tilde{B}^{z}_{2} + \tilde{B}^{z}_{3} \\-\tilde{J}^{z}_{12} + \tilde{J}^{z}_{13} - \tilde{J}^{z}_{23}}&   \tilde{J}^{x}_{23} + \tilde{J}^{y}_{23}&     \tilde{J}^{x}_{13} - \tilde{J}^{y}_{13}\\
   \tilde{J}^{x}_{13} + \tilde{J}^{y}_{13}&   \tilde{J}^{x}_{23} + \tilde{J}^{y}_{23}&\substack{\tilde{B}^{z}_{1} + \tilde{B}^{z}_{2} - \tilde{B}^{z}_{3} \\+\tilde{J}^{z}_{12} - \tilde{J}^{z}_{13} - \tilde{J}^{z}_{23}}&     \tilde{J}^{x}_{12} - \tilde{J}^{y}_{12}\\
  \tilde{J}^{x}_{23} - \tilde{J}^{y}_{23}&   \tilde{J}^{x}_{13} - \tilde{J}^{y}_{13}&   \tilde{J}^{x}_{12} - \tilde{J}^{y}_{12}& \substack{- \tilde{B}^{z}_{1} - \tilde{B}^{z}_{2} - \tilde{B}^{z}_{3}\\+\tilde{J}^{z}_{12} + \tilde{J}^{z}_{13} + \tilde{J}^{z}_{23}}\\
    \end{bmatrix}
\end{align}
\end{widetext}
\end{subequations}
\noindent We further clarify that the top-block, $\mathbf{H_{3}^{D1}}$, is controlled by parameters $\left\{ B^{z}_{i}; J^{\gamma}_{ij} \right\}$ whereas the bottom block, $\mathbf{\tilde{H}_{3}^{D1}}$, is controlled by $\left\{ \tilde{B}^{z}_{i}; \tilde{J}^{\gamma}_{ij} \right\}$.  

\section{Transformation matrix for obtaining the ion-trap parameters  $\left\{ B^{z}_{i}; J^{z}_{ij} \right\}$ and related error for a three qubit  system}\label{Diagonal-transform-3qubit}
 For the 3-qubit case illustrated here, 
 the orthogonal component 
 of the transformation matrix $[\mathbf{T}]^{T}$, can be obtained by comparing, the tensor product of two Hadamard matrices, 
    \begin{equation}
    H_{2} = \frac{1}{2}\begin{pmatrix}
        1 & 1 & 1 & 1\\
        1 & -1 & 1 & -1\\
        1 & 1 & -1 & -1\\
        1 & -1 & -1 & 1
        \end{pmatrix}
        \label{2qubit-hadamard}
    \end{equation}
and 
\begin{equation}
    [\mathbf{T}]^{T} = \begin{pmatrix}
        1  &  -1    &-1 &    1\\
        1  &  -1    & 1 &   -1\\
        1  &   1 & -1   &   -1\\ 
        \end{pmatrix}
        \label{T_top}
    \end{equation}
to arrive at  the first row vector of Eq. (\ref{2qubit-hadamard}). This row is the zero-frequency component or the average of the diagonal elements of the molecular Hamiltonian and, therefore, simply results in a uniform shift of the diagonal elements. This results in a constant shift between the eigenvalues of the molecular Hamiltonian and the Ising Hamiltonian, 
which results in no change in the dynamics, as is  clear from Figures \ref{propagation_error}, 
\ref{fig:my_label} and \ref{fig:1}.

\section{Number of degrees of control in the Ising Hamiltonian, Eq. (\ref{HIT-gen})}
\label{DoF_Ising}
For a given number of qubits, $N$, the number of ion-trap handles in Eq. (\ref{HIT-gen}) 
that control various sectors of the Hamiltonian matrix scale as 
\begin{align}
    \left\{ N + N(N-1)/2 \right\} + \left\{ N(N-1) \right\} + \left\{ 2N \right\} \rightarrow {\cal O} \left( N^2 \right),
    \label{IT-handles1}
\end{align}
Here (a) the first quantity, $\left\{ N + N(N-1)/2 \right\}$, refers to the parameters, $\left\{ B_{i}^z;  J_{ij}^z \right\}$, that control the diagonal elements of the matrix, (b) the second quanity on the left, $\left\{ N(N-1) \right\}$, refers to the parameters, $\left\{ J_{ij}^x \pm J_{ij}^y \right\}$, that control the coupling between the basis vectors inside each block, 
and (c) finally $\left\{ 2N \right\}$ refers to the parameters, $\left\{ B_{i}^x \pm \imath B_{i}^y \right\}$, that control the coupling across the sets of basis vectors created by using the odd and even raising operators described above. This characterization not only elucidates the degrees of freedom of the Ising model Hamiltonian in Eq. (\ref{HIT-gen}),
but also provides the sectored availability of these control parameters. 

At this stage there are two cases that become interesting insofar as mapping to realistic systems is concerned. In the first case, the structure of the Ising Hamiltonian is used as is, including the $\left\{ B_{i}^x \pm \imath B_{i}^y \right\}$ terms, and the number of degrees of freedom is as given above and must match the same for the problem at hand to produce an accurate map. For the second case, if the $\left\{ B_{i}^x \pm \imath B_{i}^y \right\}$ handles are eliminated, the system reduces to two separate blocks, that may be propagated independently, perhaps even on two different sets of  ion-trap architectures arranged in parallel, or Trotterized on one single ion-trap architecture. It is this second case that we consider in this paper as it allows the ability to have different Ising model parameters for the two diagonal blocks, and in this case the number of ion-trap handles become: 
\begin{align}
     2\left\{ N + N(N-1)/2 +  N(N-1) \right\}. 
    \label{IT-handles2}
\end{align}
which is, in fact, greater than the number of Ising model handles available in an $(N-1)$-qubit system when $N<17$. 
The above discussion also implies that for Hamiltonians containing $2^N$ independent terms, only approximate computation is possible. In this sense, the current paper takes a first step towards providing the necessary accuracy bounds. (Precise error bounds are provided  in Section \ref{map-diag}.)

\bibliography{PRX-refs}

\end{document}